\documentclass[preprint,12pt,authoryear]{elsarticle}

\usepackage{epsfig}

\usepackage{amssymb}
\usepackage{amsmath}
\usepackage{color}
\usepackage[T1]{fontenc}

\usepackage{booktabs}

\newcommand\nc\newcommand
\nc\be{\begin{equation}}
\nc\ee{\end{equation}}
\nc\mss{\mathrm{mss}}
\nc\demi{\frac{1}{2}}

\everymath{\displaystyle}
\def\cali#1{\cal#1\mit}
\def\abs#1{\vert #1\vert}
\journal{Remote Sensing of the Environment}
\usepackage{hyperref}
\begin{document}

\begin{frontmatter}



  \title{Revisiting the Cox and Munk wave-slope statistics using IASI observations of the sea surface }

\author[a]{Charles-Antoine Gu\'erin} \ead{Correspondence: guerin@univ-tln.fr}\author[b]{Virginie Capelle}\author[b]{Jean-Michel Hartmann}

\affiliation[a]{organization={Mediterranean Institute of Oceanography (MIO), Universit\'e de Toulon, Aix-Marseille Universit\'e, CNRS, IRD},
            addressline={Toulon},
            country={FRANCE.}}
\affiliation[b]{organization={Laboratoire de M\'et\'eorologie Dynamique/IPSL, CNRS, Ecole Polytechnique, Institut polytechnique de Paris, Sorbonne universit\'e, PSL research university},
            addressline={F-91120 PALAISEAU}, 
            country={FRANCE.}}

\begin{abstract}
  We use radiances collected from space by the Infrared Atmospheric Sounder Interferometer (IASI) when looking down at ocean surfaces during the day to remotely determine the probability distribution of wave slopes. This is achieved by using about 300 channels between 3.6 and 4.0 $\mu$m and a  physically-based approach which properly takes the contribution of the reflected solar radiation into account. Based on about 150 millions observations, the same number of wave-slope probabilities are retrieved for wind speeds (at 10 m) up to 15 m/s. We revisit and discuss the methodology proposed by Cox and Munk (CM) to derive their celebrated wave-slope probability distribution function (pdf) from photographs of the sun glitter. We propose an original and robust approach for accurate retrievals of the 7 parameters appearing in the Gram-Charlier representation of the pdf. Our results for the mean  square slopes (MSSs) are fully compatible with those of CM, and with the more recent results by Br\'eon and Henriot, but our lower uncertainties enable to point out departures from the linear wind-speed dependencies and a slight overestimation of the upwind MSS described by the linear fit of CM at moderate wind speed. Our skewness and kurtosis coefficients show clear influences of the wind speed, with a steady decrease of the former and the alongwind kurtosis coefficient being maximal at moderate wind speeds, features that CM could not point out due to the limitations of their measurements. We revisit the renormalization procedure employed by CM to obtain the complete variances from truncated pdfs and show that it imposes stringent conditions on the kurtosis coefficients that allow to determine them accurately, with wind-dependent values specific to the local sea state.  We also provide measurements of the shifted position of the most probable slope as well as a demonstration of a qualitative change of regime in the updown wind asymmetry of the wave-slope probability when the wind speed increases. 



\end{abstract}







\end{frontmatter}


\section{Introduction}
\label{intro}
The wave-slope probability distribution function (pdf), which statistically describes the tilts of the facets at the sea surface, has many implications in geosciences. Indeed, since it determines the area of the water-air interface, it must be known for  calculations of the energy and mass exchanges between the oceans and the atmosphere involved in  meteorological and climate studies.  Knowing the pdf is also important for various remote sensing issues including retrievals of the sea-surface temperature [e.g. \citep{MerchantSST,Capellenighttime,Capelledaytime}], of aerosols and clouds [e.g. \citep{aerosols1,aerosols2,aerosols3,clouds1,clouds2}], of salinity [e.g. \citep{salinity2,salinity1}], of oil spills and slicks [e.g. \citep{oil3,oil2,oil1}], of the ocean color [e.g. \citep{color2,color1,color3}],  and of the wind speed [e.g. \citep{Breon,Zhang2018,Nelson}]. In addition to these examples, recall that the wave-slope statistics also enters in the physical parametrization of spectral wave models through its second-order moments [e.g. \citep{stopaOM2016}]. and that the key role that its plays is proven by the very large number of papers citing the seminal study of \cite{Cox54} who proposed its first accurate parameterization.

Due to its importance, the wave-slope pdf has been the subject of numerous investigations. As far as field observations of the sea surface are concerned, the pioneer study was carried seventy years ago by \cite{Cox54,Cox54b} [see also \citep{Cox56} for a documented report on the experiment], from now on denoted as CM, who used black-and-white photographs of the sun glitter taken from an aircraft. Despite the limits of the technique used and the very limited number of observations (25 for clean sea), this groundbreaking investigation enabled to parameterize the influences of the wind speed on the up-, -down, and cross-wind slope probabilities, which were found different. The parameters of the proposed Gram-Charlier distribution, i.e.: The mean square slopes, and the skewness and kurtosis (peakedness) coefficients, were described through linear (or constant) functions of the wind speed (up to 14 m/s) at 12.5 meters above the surface. The main limitations of this seminal work are that the measurements only provided the relative variations of the intensity of the sun glitter, due to lack of absolute calibration, and that no information was obtained for large wave-tilt angles $\theta_w$ for which the sun glint was drowned in the background radiations. The maximum available value of $\theta_w$ was between $15^\circ$ and $25^\circ$ depending on wind speed. Extrapolations thus had to be made and the wave-slope pdf needed to be normalized. Note that \cite{Wu} later on reanalyzed the data of CM and found that the influence of the wind on the isotropic distribution is, for  speeds below (resp. above) 7 m/s, less (resp. more) pronounced than originally found. A new parameterization was proposed, based on a two-branch logarithmic function leading to probabilities differing from the former ones by several tens of \% in some cases. Since then,  several other field investigations have been carried out, which are presented below in chronological order. Note that in all cases the information on the wind direction and speed was provided by collocated (in time and space) measurements.  \cite{Hughes} determined the wave-slope pdf from the refraction, measured by a camera carried by a crane 10 m above the bow of a ship, of the beam emitted by an underwater He-Ne laser. The results, obtained for 9 wind speeds between 3.6 and 8.2 m/s and wave tilts up to 22$^{\circ}$, are in substantial agreement with those of CM. Then, \cite{Mermelstein} obtained slope probabilities, for wind speeds up to 20 m/s, from the wave-height power spectral density deduced from radar observations by \cite{Donelan}. With respect to that of CM, their pdf is significantly broader in both the up- and cross-wind directions, with less asymmetry for wind speeds above 5 m/s. A few years later, by using their laser-glint measurements complemented by those of \cite{Hwang},  \cite{shaw} introduced a correction to the pdf of CM in order to take the influence of the degree of atmospheric stability near the sea surface into account. This was achieved by multiplying the original mean square slopes by a function depending on the Richardson number $R_i$. The results are identical to those of CM for $R_i$=0.15 but differ by more than +40 \% for neutral and unstable states ($R_i \leq 0.$), while a -35 \% correction applies for $R_i \geq 0.27$. Then, \cite{Ebuchi} used $30$ millions (0.55-0.75 $\mu$m) images from the Visible Infrared Spin-Scan Radiometer (VISSR). The pdf obtained in the 0-10 m/s wind-speed range is quite close to that of CM for the crosswind component, but very different (narrower) for the alongwind one, which leads to a much less pronounced (and inverse)  up- to cross-wind asymmetry. These differences were attributed to the fact that the observations of CM were made under conditions of growing waves, a questionable explanation in view of the results of \cite{Breon} discussed below. Again, no probability was derived  for steep waves since the investigated tilt angles are all smaller than  30$^{\circ}$. \cite{Breon}, from now on denoted as BH, carried an investigation using a dataset of $6$ millions radiances at 0.865 $\mu$m collected by the Polarization and Directionality of the Earth's Reflectances (POLDER). This led to results, for wind speeds up to 15 m/s, which confirm those of CM, except for some skewness coefficients which were found to vary non-linearly with the wind speed [also see the discussion in \cite{Munk}], but suffer from similar limitations with $\theta_w \lesssim 22^{\circ}$ and the use of extrapolations and of a normalization. In \cite{Ross}, observations between 0.444 and 0.864 $\mu$m  where used to retrieve slope variances (but no analytical model was derived from them) for the 0-14 m/s wind-speed range. The results follow the general trend established by CM, but point out a non-linear  trend with wind speed for the crosswind direction, as well as, in the lower wind-speed regime, for the upwind one.  Finally, the most recent study was carried \citep{Lenain} by analysing the glint of an airborne laser. Information was retrieved for wave-tilt angles up to $26^{\circ}$ in the 2-14 m/s wind-speed range, leading to up- and cross-wind  mean square slopes in good agreement with those of CM, but to differences for the skewness and peakedness. Since observations in the optical domain are much more sensitive to steep and small wave facets that do probings at long wavelengths, no review on this last topic is made here. Let us however mention that the wave-slope statistics estimated from microwave radar observations, e.g. \citep{jackson1992,Vandemark,boisot2015,SWH1,Chen},  generally fall in between those of CM for a clean and a slick sea surface, because the information on facets with dimensions smaller than the wavelength is lost.
For completeness, recall that, besides the above mentioned studies, other ones are limited to tests of previously proposed pdfs, made using fields measurements.  \cite{Zhang2010} compared sun-glint observations at 0.859, 1.24, and 2.13 $\mu$m from the Moderate Resolution Imaging Spectroradiometer (MODIS) with many of the above mentioned parameterizations. They concluded that the facet models of CM and BH lead to the best predictions. The CM parameterization of the pdf was also tested in numerous  field investigations in a broad range of wavelengths, e.g.  \citep{Haimbach,Su,Vandemark,Gatebe,Hauser,Liang,Yurovskaya,Zhang2018}. In most cases, a globally satisfactory agreement was obtained provided that, for observations at wavelengths exceeding the dimensions of some wave facets, the associated filtering effect is taken into account. In conclusion, the bibliography analysis shows that there are large inconsistencies between some of the proposed pdfs, as illustrated by Fig. 4 of \cite{Zhang2010}, where the results obtained using the parameterizations of \cite{Mermelstein} and \cite{Ebuchi} differ by a factor of over 3, those from \cite {Cox54,Cox54b} and \cite{Breon} being in-between. Although the findings of \cite{Breon}, \cite{Zhang2010}, \cite{Ross}, \cite{Zhang2018}, and \cite{Liang} are in favor of the of CM, some uncertainties remain on the latter. The major one results from the  use of extrapolations and of a renormalization. In addition, the issue concerning the linear dependence of the mean square slopes on wind speed remains open, some studies having suggested a different behavior [e.g. \citep{Ross,Wu}]. Furthermore, the influence of the wind on the kurtosis and skewness coefficients remains very uncertain. These elements indicate that further investigations are still needed and that using, for this purpose, observations provided by wavelengths and an instrument different from those so far used is desirable.

In \cite{Capelledaytime}, daytime spectra recorded by the Infrared Atmospheric Sounder Interferometer (IASI) satellite instrument \citep{IASI} were used to retrieve  sea-surface temperatures (SSTs) from radiances at various channels between 3.6 and 4.0 $\mu$m. This required to take the contribution of solar photons reflected by the surface into account. This was achieved, for each observation, by simultaneously determining the SST and a parameter (directly related to the wave-slope probability) scaling the computed solar contribution through fits of the IASI data using a physically-based forward model. Note that \cite{Capelledaytime} focused on the SST and on the validation of the methodology proposed for its retrieval, which imposed the use of a limited set of  observations: Those for which a collocated in-situ measurement of the temperature from a drifter is available. The consequently relatively small number ($10^5$) of wave-slope probabilities obtained forbid any reliable investigation of the pdf, but the comparisons made with the parameterizations of CM and BH demonstrated a good agreement.

In the present study, now that the excellent agreement between the IASI-obtained SSTs and in-situ measurements  \citep{Capelledaytime} has given us a high confidence in the robustness of the methodology used, we investigate the wave-slope probabilities. This is made possible by the treatment of a much larger set (about 150 million) of data, thanks to the release of the collocation-with-drifters constraint. The remainder of this paper starts, in Sec. 2, with a presentation the IASI spectra used. The forward model, its input data, and the procedure applied to fit the measured radiances are then described in Sec. 3, the main characteristics of the  IASI-retrieved dataset of wave-slope probabilities being the subject of Sec. 4.  Section 5 then details how  the latter were analyzed, which is done after recalling the approach used by CM. The results obtained are then presented and discussed in Sec. 6, where comparisons with previous studies are made.

\section{The IASI spectra}
	The satellite-based Infrared Atmospheric Sounding Interferometer (IASI) collects spectra of the radiation coming from below \citep{IASI}. It provides extremely well-calibrated radiances at 8461 equally-spaced points from 645 to 2760 cm$^{-1}$ (15.5 to 3.6 $\mu$m), with a spectral resolution of 0.50 cm$^{-1}$  after apodization (for the Level-1c spectra used here). The instantaneous field of view leads to a ground resolution of 12 km at nadir over a swath of about 2200 km (+/-50 $^{\circ}$). Three successive generations of IASI fly onboard the Metop-A, -B, and -C series of polar-orbiting satellites, with Level-1c data available since July 2007, February 2013, and July 2019, respectively. All three satellites have a Local Time of descending orbit node of 9:30 AM (+/-15 mn) at the Equator. The spectral and  radiometric stabilities of IASI are continuously controlled by the Centre National d'Etudes Spatiales ensuring an absolute accuracy better than 0.5 K at 280 K on brightness temperatures and a relative one lower than 2 ppm on wave numbers \citep{Blumstein2004,Blumstein2007} . 
	
	In the present study, a set of about 150 million IASI daytime observations was used. The spectra, recorded from Metop-A (2008-2021) and -B (2015-2019) were selected, as done previously \citep{Capellenighttime,Capelledaytime}, using the following criteria: (i) Only cloud- and aerosol-free data were retained, selected by using criteria on brightness temperatures from IASI and the collocated AVHRR instrument. (iii) We only kept observations close to nadir ($\leq$ 30$^{\circ}$) in order to minimize potential analysis biases resulting from the dependence \citep{Masuda88} of the sea-surface emissivity on the wind speed. (iii) Only those spectra for which the value of the wind speed at 10 m above the observed spot is smaller than 15 m/s were retained for the analysis due to the increasing potential contamination by whitecaps \citep{Anguelova,Anguelova2} and the reduced number of observations at large wind speeds. 

\begin{figure}[htbp]\centering
	  \includegraphics[scale=0.40]{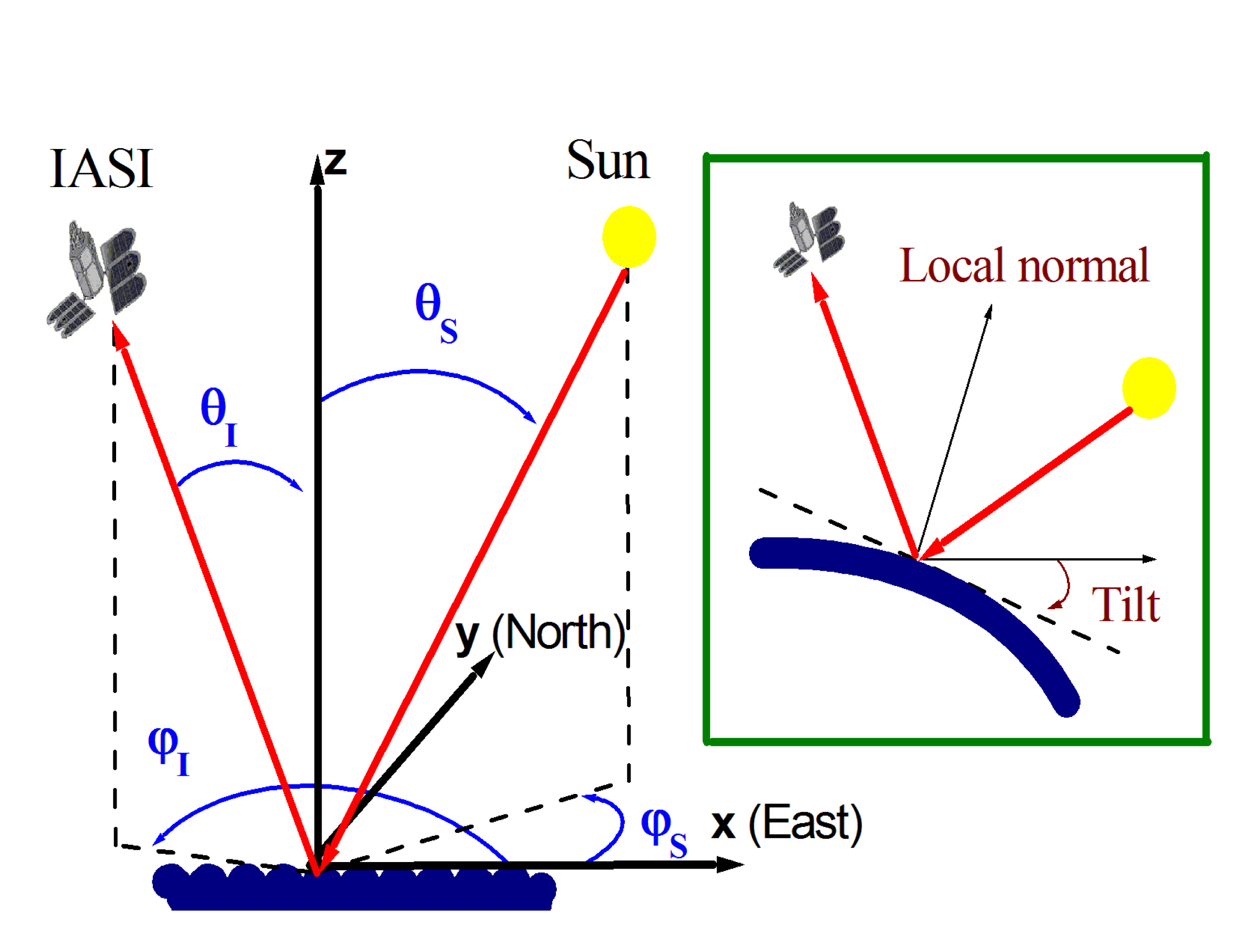}
\caption{Observation geometry. The z axis is along the local vertical, x points toward the East and y toward the North. The insert enlightens the fact that, the refection of light on the surface being specular, only an ad-hoc wave slope redirects the solar photons toward the instrument.}
\label{fig:IASI}
\end{figure}

\begin{figure}[htbp]\centering
	  \includegraphics[scale=0.4,angle=-90]{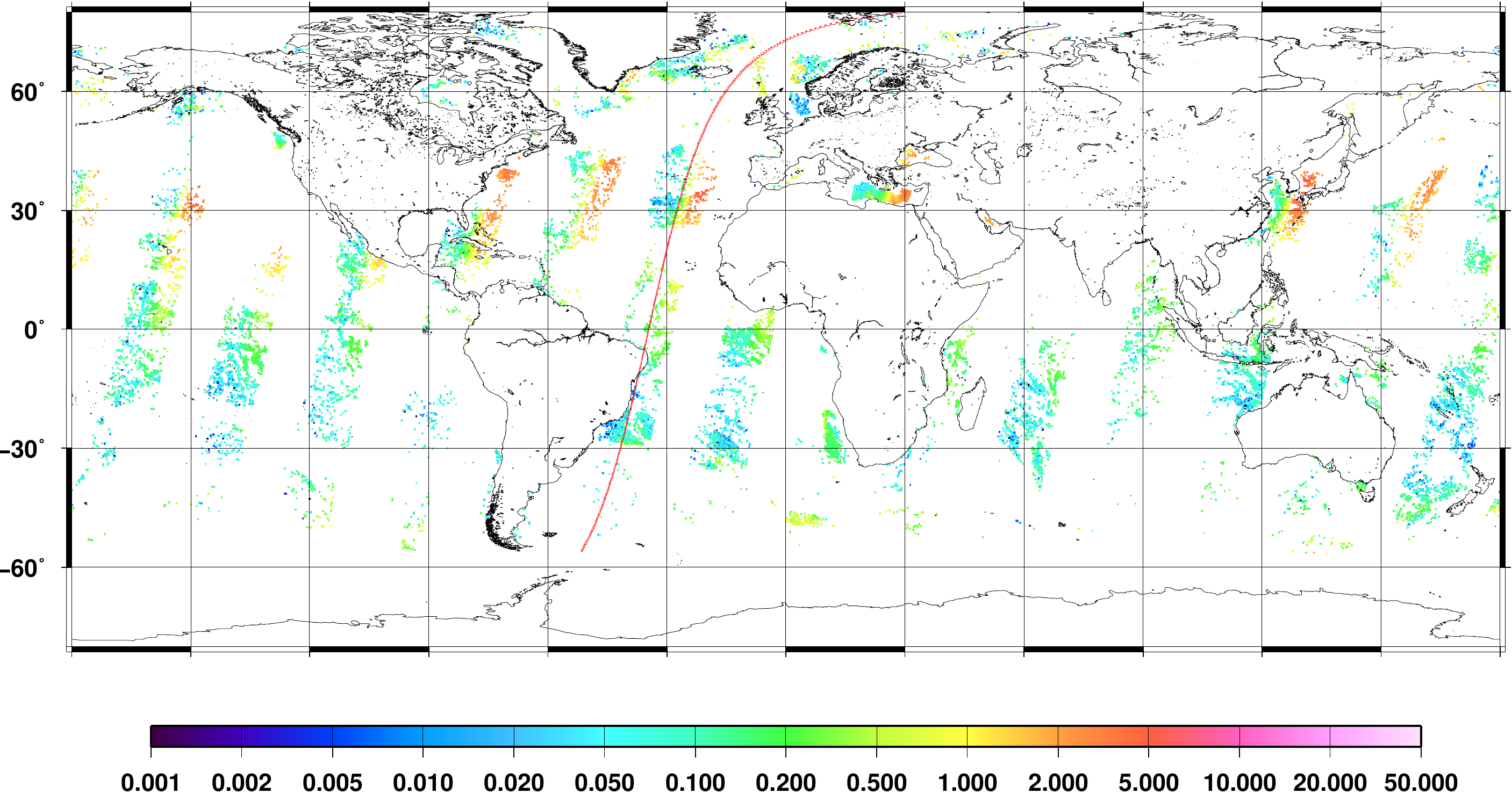}
\caption{Wave-slope probability (logarithmic scale) retrieved, as explained in Sec. 3, from the spectra collected by IASI along the descending orbits of 16 July 2018. The red line on the central orbit indicates the center of the swath (nadir pointing).}
\label{fig:mapproba}
\end{figure}

In many observations, the solar contribution to the measured radiances (and the associated wave-slope probability) is small, except when the instrument points close to the sun glint. This latter case corresponds, for the geometry of Fig. \ref{fig:IASI} and with the convention that all zenith angles are positive, to small values of both $|{\theta_I}-{\theta_S}|$ and $|{\varphi_I}-{\varphi_S}$-180$^{\circ}|$. 
The daytime descending orbits being close to the North-South direction, the swath roughly follows an East-West line. $\varphi_I$ is thus close to 0$^{\circ}$ for a Westward pointing and to 180$^{\circ}$ for and Eastward one. All observations being made at 9:30 AM Local Time (at the Equator), the sun is at the East and a significant solar contribution can only be collected on this side of the swath and when $\varphi_S$ is close to 0$^{\circ}$ while $\theta_S$ is not to far outside the 0-30$^{\circ}$ range of retained $\theta_I$ values. During the summer, large wave-slope probabilities, which correspond to weakly tilted wave facets, will thus be observed at moderate North latitudes only, because the sun inclination to the North is too large (i.e.  ${\varphi_S}$ is too far above 0$^{\circ}$) in the Southern hemisphere, the situation being reversed in the winter. This is confirmed by the retrieved wave-slope probability map displayed in Fig. \ref{fig:mapproba}.

Using IASI radiances, which are all for wavelengths greater than 3.6 $\mu$m,  for retrieving the wave-slope pdf from the information brought by the reflected solar light may \textit{a priori} seem a not optimal choice when compared to observations made in the visible or near infrared [as done in e.g. \cite{Cox54,Cox54b,Breon,Ebuchi,Zhang2010,Ross}]. Indeed, at such significantly shorter wavelengths, not only is the sun radiance larger, but the thermal emissions of the ocean and atmosphere can be neglected. However, using IASI data has several noticeable advantages. The first is that they are extremely well calibrated, which enables to retrieve absolute values of the pdf and thus to avoid the use of any normalization procedure. The second is that the calibration is, for a given instrument, remarkably stable over time, and that the consistency between the three generations of IASI onboard Metop-A, -B, and -C is also extremely good. These characteristics, demonstrated in \cite{Capellenighttime}, enable to use long time series and several instruments,  thus providing a very large dataset of wave-slope probability values. Finally, the potential biases introduced by un-rejected observations contaminated by aerosols or due to errors in the modeling of light scattering in the atmosphere are reduced. Indeed, the involved cross-sections are much smaller around 4 $\mu$m when compared with the visible region. In the mid-infrared, only the aerosols composed of the largest particles (such as mineral dust or sea-salt) can significantly contribute to the collected radiances (by scattering or absorbing the photons), while aerosols typical of pollution or biomass burning have a negligible impact \citep{Pierangelo2013}. Mineral dust have the largest influence, but they can be detected [e.g. \citep{Capelle2018}] and the contaminated observations can therefore be disregarded  as done in the present study.   Concerning the sea-salt aerosols, they are principally located at the bottom of the planetary boundary layer, a region of small impact on the radiances collected in the spectral range considered here. \textcolor{black}{Finally, the influence of the eventual presence of white caps is also reduced, since their reflectivity is, with respect to that of a clean sea surface, significantly less increased around 4 $\mu$m \citep{Salisbury1,Salisbury2} than it is \citep{Frouin, Dierssen} at the short infrared [used by \cite{Breon}] and visible [used by \cite{Cox54}] wavelengths}. 

\section{Slope probabilities retrieval procedure}
  	Since details on the theoretical model and data used can be found a previous study \citep{Capelledaytime} and are recalled in Appendix A, only the main elements are given below. 

        \subsection{Forward model}
For the observing geometry schematized in Fig. \ref{fig:IASI}, the theoretical expression of the IASI radiance at wave number $\sigma$, which disregards any light scattering in the atmosphere, is: 
\begin{equation}
 \begin{split}
I_{theo}(\sigma,\theta_I,\varphi_I,\theta_S,\varphi_S)=I_{surf}(\sigma,\theta_I,\mathbf{T_{surf}} )
 +I_{atm}^+(\sigma,\theta_I)\\
+I_{atm}^-(\sigma,\theta_I)+ \mathbf{ P_{w}} (\theta_I,\varphi_I,\theta_S,\varphi_S)I_{sun}(\sigma,I,\theta_I,\varphi_I,\theta_S,\varphi_S).
 \end{split}
 \label{eqIASI}
\end{equation}
The first term , $I_{surf}$,  is the contribution of the surface emission which can be expressed in terms of the sea directional emissivity and skin temperature (SST) ($T_{surf}$), and of the atmospheric transmission from the surface to the instrument. The second term, $I_{atm}^+$,  which represents the emission of the atmospheric column between the surface and the instrument, can be computed provided that the vertical profiles of temperature, pressure and volume mixing ratios of absorbing species are known. The third term, $I_{atm}^-$, which results from the emission of the atmosphere toward the surface that is reflected and transmitted up to the instrument, depends on the same variables, complemented by the reflectivity of the surface. Finally, the last term describes the contribution of reflected solar photons. As shown in Appendix A, it  involves the out-of-the-atmosphere sun radiance, atmospheric transmission, bi-directional reflectivity of the sea surface, and probability $P_{w}(...)$ for waves to have a slope enabling solar photons coming along the  $(\theta_S,\varphi_S)$ direction to be redirected toward the instrument, along  $(\theta_I,\varphi_I)$, after a specular reflection at the air-water interface (see insert in Fig. \ref{fig:IASI}). As highlighted by the use of bold characters in Eq. (\ref{eqIASI}), $T_{surf}$ and $P_w$ are the only unknown quantities of the model, that will be retrieved (see Sec. 3.3) from fits of IASI-measured spectra. Indeed, all the other parameters needed for the calculation of the radiances (atmospheric state, sea-surface emissivity and reflectivity, absorption coefficients of absorbing gases, solar radiance) can be obtained from independent sources, as explained below.

\subsection{Input data used}
	 The index of refraction for pure water of \cite{Downing} was used, after a shifting of +4 cm$^{-1}$ and an increase of the real part by +6 10$^{-3}$ \citep{Friedman} in order to take salinity effects into account, to compute the sea surface emissivity  as done by \cite{Masuda88}, and the bi-directional reflectivity  as explained in Appendix A. In the 3.6 to 4.0 $\mu$m region used for the retrievals, the extraterrestrial solar radiance was represented by a Planck function with a temperature of 5657 K, consistent with \cite{Platnick}. The vertical profiles of temperature, pressure, and volume mixing ratios of the absorbing species of each atmosphere observed by IASI were obtained \citep{Capelledaytime} using a proximity recognition within the Thermodynamic Initial Guess Retrieval (TIGR) database \citep{Chedin,Chevallier} for a selection of 8 properly chosen channels. The spectroscopic line parameters of the absorbing gases, needed for the computation of the atmospheric spectra, were taken from the GEISA database \citep{GEISA} [\href{https://geisa.aeris-data.fr/}{aeris}]. They were complemented by the MT\_CKD (version 3.1) water-vapor continua \citep{Mlawer} [\href{http://rtweb.aer.com/continuum\_frame.html}{MTCKD}], and the collision-induced absorption by N$_2$ and the CO$_2$ line-wings corrections were computed as in \cite{Hartmann18}. All these data were used as inputs to the 4AOP radiative-transfer code \citep{Scott,Cheruy} 
	[https://4aop.aeris-data.fr/] for predictions of the atmospheric emissions and transmissions. 

\subsection{Retrieval procedure and its validation}
For the treatment of the observations, the spectral points of locally minimum absorption selected \citep{Capelledaytime} in two windows around 3.7 and 4.0 $\mu$m have been retained (107 points between 2480 and 2528 cm$^{-1}$ and 185 points from 2594 to 2760 cm$^{-1}$). The analysis then consists, for each IASI-measured  spectrum $I_{IASI}$, in its least-square fit by the theoretical expression in Eq. (\ref{eqIASI}). In other words, the quantity
\begin{equation}
\begin{split}
\Sigma_{\sigma_k}\left[I_{IASI}(\sigma_k,\theta_I,\varphi_I,\theta_S,\varphi_S)
-I_{theo}(\sigma_k,\theta_I,\varphi_I,\theta_S,\varphi_S, T_{surf},P_w) \right]^2,
\end{split}
\end{equation}
where the sums extends over the 292 retained channels $\sigma_k$,
    is minimized by simultaneously floating  the $T_{surf}$ and $P_{w}$. This fit has little ambiguity thanks to the small correlation between the adjusted parameters brought by the use of a spectral range over which the ratio of the radiances coming from the sea-surface emission and from the sun varies by a factor of over 3.
    
\begin{figure}[htbp]\centering
	  \includegraphics[scale=0.4,angle=0]{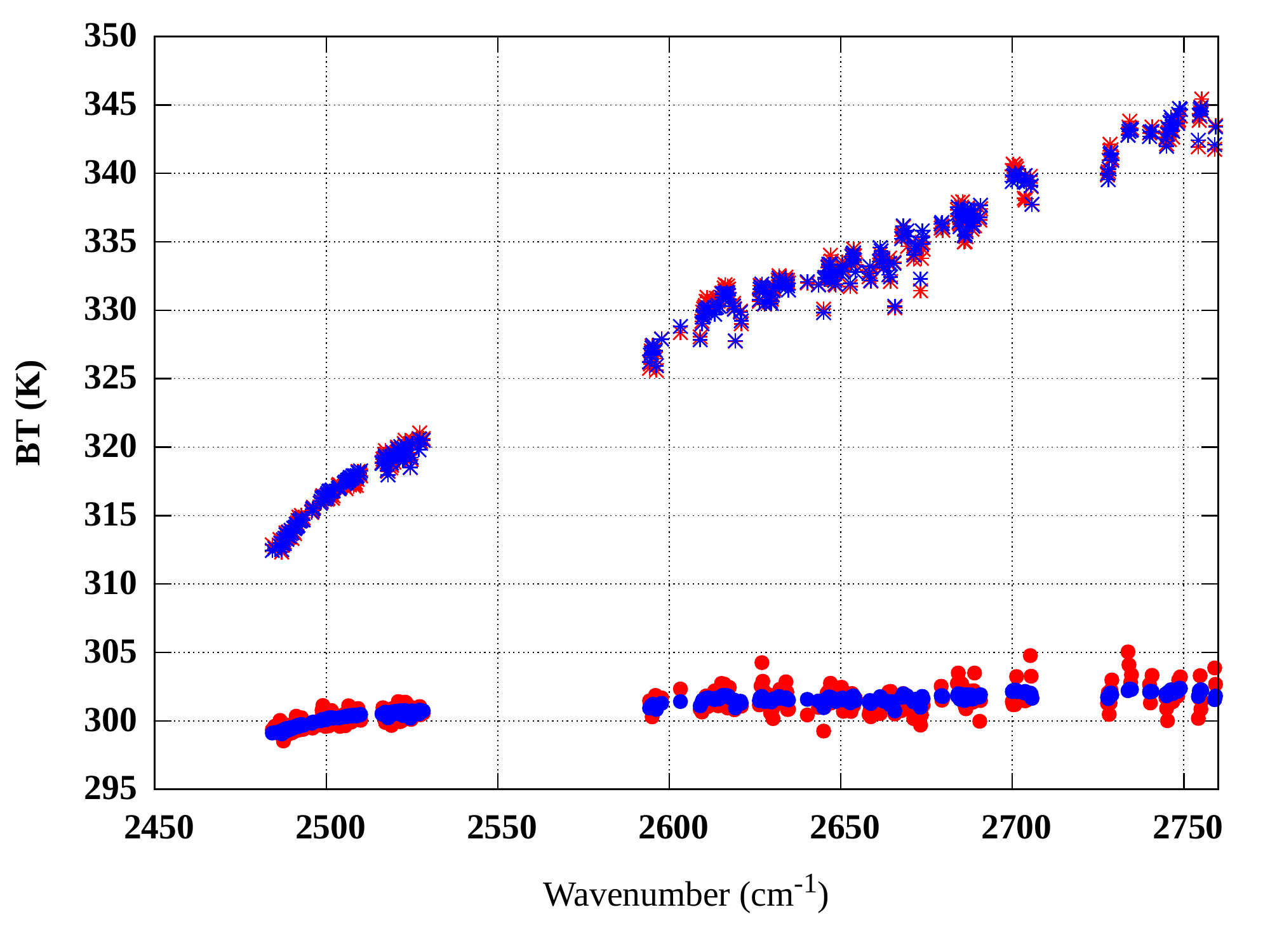}
\caption{Observed (red symbols) and calculated+fitted (blue symbols) brightness temperatures for observations including a large (bright case: upper series of points) and very small (dark case: lower series of points) solar contribution.}
\label{fig:Fit_IASI}
\end{figure}

    Typical comparisons between measured and adjusted brightness-temperatures  for observations with a small and a large solar contribution are displayed in Fig. \ref{fig:Fit_IASI}. Recall that the brightness temperature $BT(\sigma)$ associated with a radiance $I(\sigma)$ is such that the blackbody (Planck) function at wave number $\sigma$ and temperature $BT(\sigma)$ is equal to $I(\sigma)$. As can be seen, the agreement is very good and the effect of the sun, which can lead to a significant increase of the BTs with wave number, is nicely taken into account, demonstrating the qualities of the forward model and fit. As a result, the statistical uncertainty on the probability $P_w$ provided by the adjustment is very small, below 1\% for the most probable slopes thanks to a significant solar contribution to the observed radiances [we found that an error was made in the evaluation of the uncertainty on the $A$ parameter made in \cite{Capelledaytime}, leading to strongly overestimated values of $\Delta A$. This led us to wrongly attribute the observe scatter of the results to uncertainties, while a large part of it was likely due to the influence of scene parameters other than the wind, as discussed in Sec. 6.3]. Note that the differences between the BTs in the dark and bright observations of Fig. \ref{fig:Fit_IASI} correspond, for the former, to an increase, due to the sun, of the  radiance at the shortest wavelength by a factor of over 5.  Finally recall that the robustness of the procedure described above was  validated \citep{Capelledaytime} by successfully comparing the retrieved SSTs with collocated in-situ temperature measurements from drifters. Indeed, this showed, on average and after correction of the cool-skin effect \citep{Fairall}, an agreement within typically 0.01 K. In addition, the preliminary comparisons \citep{Capelledaytime}  between the fitted wave-slope probabilities and values obtained using the parameterizations of \cite{Cox54} and \cite{Breon} demonstrated a good consistency. 

\section{The IASI-retrieved wave-slope probabilities dataset}\label{sec:IASI}

As demonstrated by \cite{Cox54,Cox54b}, the analysis of wave-slope probabilities and the building-up of a pdf require knowledge of the wind speed and direction. This information could \textit{a priori} be provided by the Advanced SCATterometer [ASCAT, see \citep{ASCAT} and \href{https://www.eumetsat.int/ascat}{ASCAT}] which flies together with IASI onboard the Metop platforms. However the problem is that ASCAT only looks from 336 to 886 km away from nadir, for zenith angles in the direction perpendicular to the orbit within the +/-[$25^\circ$,$53^\circ$] range. Since we only kept (see Sec. 2) observations for $-30^\circ\leq\theta_I\leq+30^\circ$, imposing a collocation with ASCAT limits the scanned $\theta_I$ range to +/-[$25^\circ$,$30^\circ$]. This would significantly reduce the number of retrieved wave-slope probabilities, thus affecting the reliability of the obtained pdf. We therefore resorted to the wind information from the European Centre for Medium-range Weather Forcast (ECMWF) ERA5 reanalysis dataset \citep{ERA5} which provides \citep{ERA5data} world-wide values (at 10 m above the surface) on hourly and $0.25^\circ\times0.25^\circ$ spatial-grid bases. This is not optimal since these data come from a model and not from measurements. In order to quantify the possible errors, we used all  measurements from ASCAT over oceans for January, April, July and November 2019, and compared them with the corresponding (in space and time) ERA5 data. This shows that the mean errors on the wind speed remain below 1 m/s with typical RMS values after correction for the mean bias of about 1 m/s (for the 3-15 m/s range considered in this study). For the wind direction, the mean difference between ASCAT and ERA5 is within $\pm 4^\circ$ (with a change of sign when switching from the North to the South Hemisphere) with a RMS (after correction for the bias) which decreases with wind speed, from about $20^\circ$ for 3 m/s down to $6^\circ$ for 15 m/s. Note that these findings, which show that the errors are moderate, are consistent with those of \cite{Belmonte} and \cite{ERA5wind}.

For the analysis of the IASI-retrieved probabilities, the relevant variables are the wave slopes along and perpendicular to the wind direction. Following many previous studies we retain for the x axis the upwind direction. This corresponds to horizontally rotating the frame of Fig. \ref{fig:IASI} by  -$\varphi_W$, where $\varphi_W$ is the azimuth angle between the upwind and East directions. Then, the up- and cross-wind slopes, $s_u$ and $s_c$, are given by:
\begin{equation}
s_u=dz/dx=-\frac{\sin(\theta_S)\cos(\varphi_S-\varphi_W)+ \sin(\theta_I)\cos(\varphi_I-\varphi_W)}{\cos(\theta_S)+\cos(\theta_I)}
\end{equation}
\begin{equation}
s_c=dz/dy=-\frac{\sin(\theta_S)\sin(\varphi_S-\varphi_W)+ \sin(\theta_I)\sin(\varphi_I-\varphi_W)}{\cos(\theta_S)+\cos(\theta_I)}
\end{equation}

\begin{figure}
   \centering
    \includegraphics[scale=0.3]{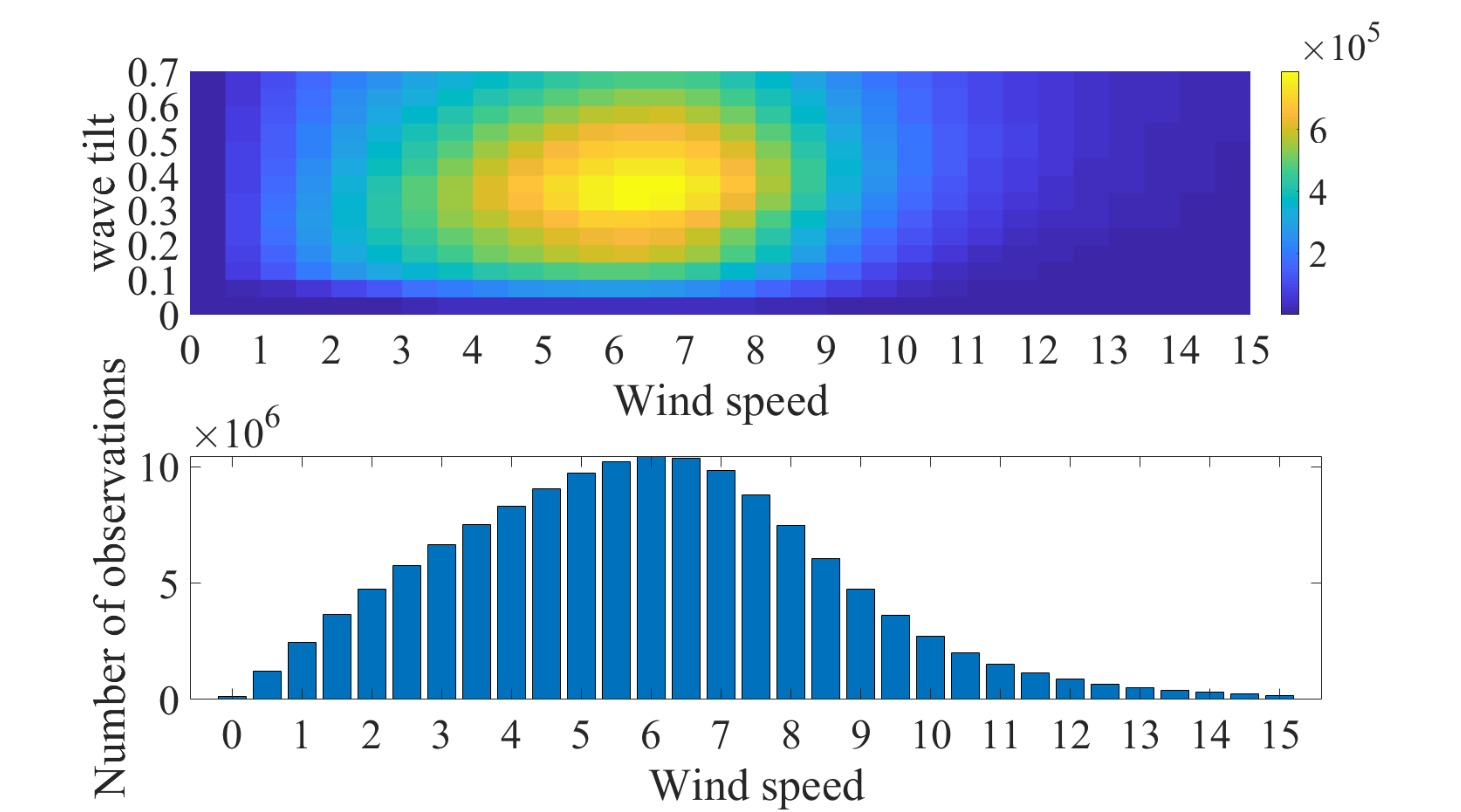}
   \caption{Number of observations  classified by wind speeds and wave tilt values (top panel) and by wind speed only (lower panel).}
   \label{fig:Ndata}
\end{figure}

The treatment of all the selected IASI spectra provided about 150 million wave-slope probabilities, with, for each of them, values (see above) of the wind speed (between 0 and 15 m/s) and direction. Figure \ref{fig:Ndata} presents the number of data points versus the wind speed (with 0.5 m/s bins) and the wave-tilt  $(s^2_c+s^2_u)^{1/2}=\mathrm{tg}(\theta_{w})$. As can be seen, the largest numbers of observations are obtained for moderate winds (5-8 m/s), which are as well known the most frequent, and intermediate wave slopes ($0.3-0.4$,  corresponding to $\theta_{w}$ between $17^{\circ}$ and $22^{\circ}$). The very small tilts are only rarely observed due to the constraints on $\theta_S$ and $\varphi_S$ imposed by the position of the sun at 9:30 AM LT and those on $\theta_I$ and $\varphi_I$ resulting from the IASI observation geometry.

\begin{figure}\centering
    \hspace{-1cm}\includegraphics[scale=0.23]{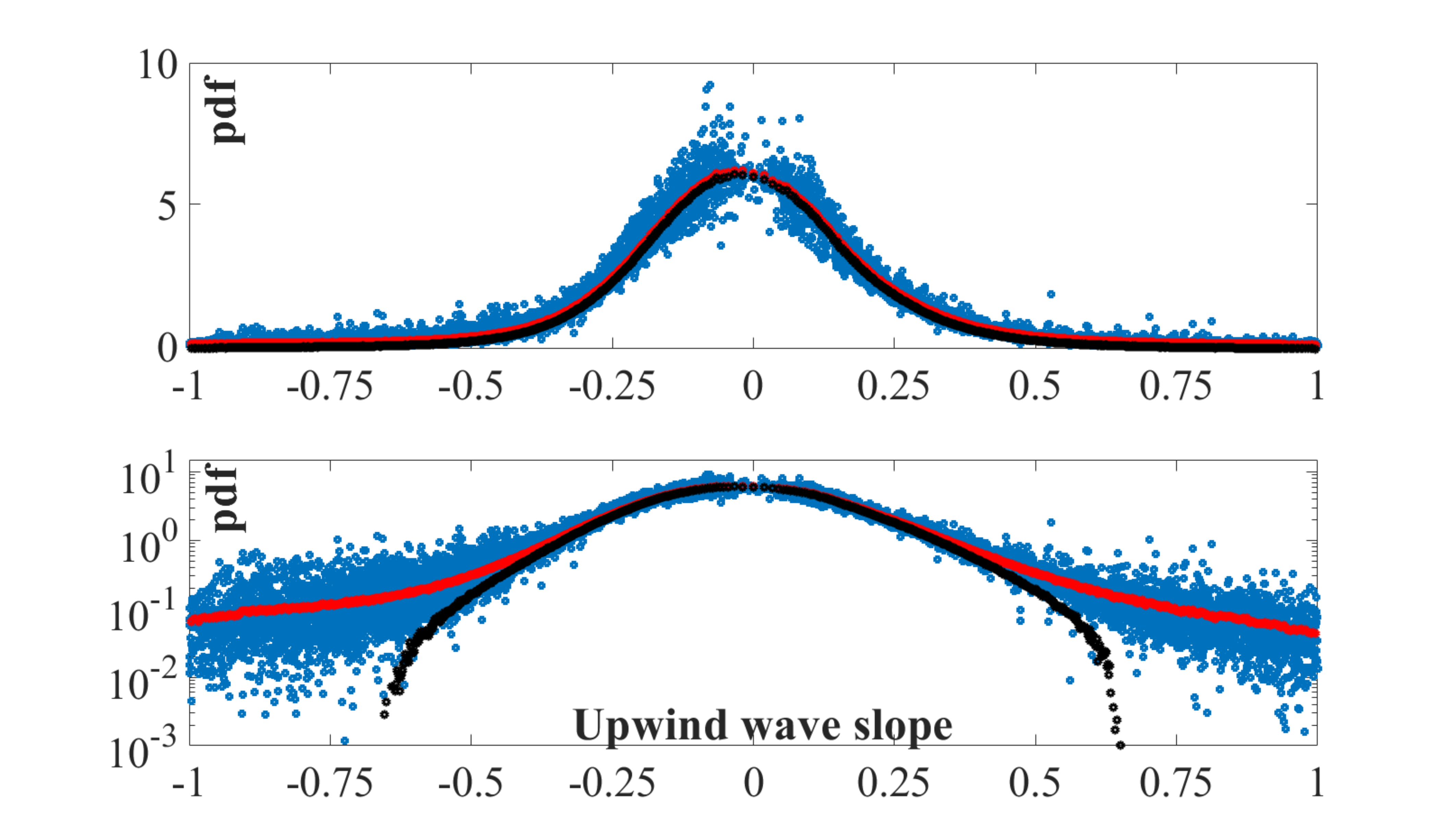}
   \caption{Measured probabilities (blue dots) of alongwind slopes in the principal section ($s_c=0$) for a wind speed of 10 m/s displayed with linear (top panel) and logarithmic (lower panel) scales. The least-square fit with 3 Fourier azimuthal harmonics [Eq. (\ref{azimfit}) is shown by the red lines. The black dotted lines is the same fit after correction for the bias.} 
   \label{fig:pdf7up}
\end{figure}

Looking at the dependence of the IASI retrieved probabilities $p(s_u,s_c,U)$ for given wind speeds $U$ shows that they are slightly biased in the tail. Indeed, as exemplified by Fig. \ref{fig:pdf7up}, the probability tends toward a small but non-zero constant value for large slopes and does not keep decreasing with increasing tilt, as it should do. The origin of this bias, which can be considered as a background floor to be rejected, is at this stage not completely established. \textcolor{black}{We hypothetize several, possibly concurrent, causes for it: a) An insufficient accuracy in the estimation of small probabilities through the above described retrieval procedure; b) The contamination of the tail of the distribution by rare but intense reflections due to near-breaking waves; c) The contribution of whitecaps and foam patches; d) Other parasitic contributions not taken into account in the model, such as aerosols and thin clouds that have not been rejected by our filtering procedure. Note that the bias has been found increasing with wind speed, which supports the hypotheses b) and/or c) but does not rule out other reasons at this stage. The very likely influence of breaking and whitecaps will be further discussed in Section \ref{discussion}.}

In order to  correct for this problem, we used the fact that integrating $p(s_u,s_c,U)$ over $s_u$ and $s_c$ should lead to unity, the idea being to subtract a constant $b(U)$ to $p(s_u,s_c,U)$ in order to fulfill this condition. However, since the integration requires to interpolate among the measured values, a fit of the latter must first be made. For this purpose, it is convenient to consider the probabilities in polar coordinates, that is $p(s,\theta)$ with $s=(s_u^2+s_c^2)^{1/2}$ and  $\theta=\mathrm{atan}(s_c/s_u)$ measured anticlockwise with respect to the alongwind axis (thus $\theta=0/\pi$ for the upwind/downwind directions while $\theta=\pm \pi/2$ for the crosswind ones). Then, for each wind speed, $p(s,\theta)$ can be expressed as an azimuthal series:
\begin{equation}
  p(s,\theta)=a_0(s)+a_1(s)\cos\theta+a_2(s)\cos(2\theta),
  \label{azimfit}
  \end{equation}
    with $s>0$ and $-\pi<\theta\leq \pi$, where we limit the expansion to order 2 since this is sufficient to account for both the upwind-crosswind and upwind-downwind asymmetries. The up- ($s_u>0,s_c=0$), down- ($s_u<0,s_c=0$) and cross-wind ($s_u=0$) probabilities are then:
    \begin{equation}
  \begin{split}
    p_{up}(s)&=a_0(s)+a_1(s)+a_2(s),\\
    p_{do}(s)&=a_0(s)-a_1(s)+a_2(s),\\
    p_{cr}(s)&=a_0(s)-a_2(s),
  \end{split}
      \label{pup}
    \end{equation}
    the omnidirectional (or isotropic component) value being $p_{omni}(s)=a_0(s)$.

\begin{figure}\centering
    \includegraphics[scale=0.25]{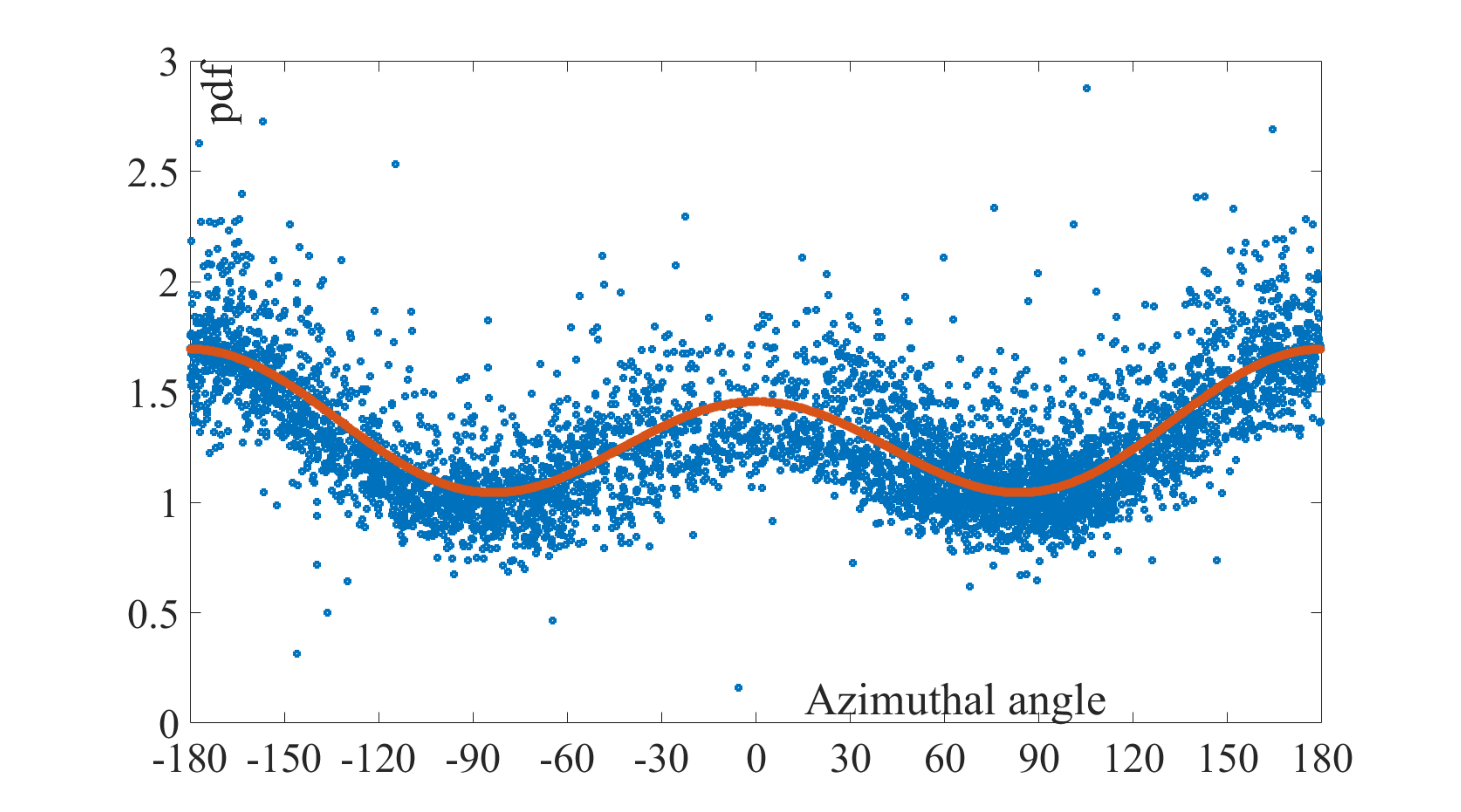}
   \caption{Azimuthal variations of the measured probabilities (blue dots) together with their least-square fit (red solid lines) using Eq. (\ref{azimfit}) for a wind speed of 10 m/s and a slope $s_i$=0.29 ($16.3^\circ$).}
   \label{fig:exemplefit}
\end{figure}

    In order to determine the values of $a_{i=0,1,2}(s)$, we first sorted the data by wind speed using a $0.5$ m/s step and centered bins of width $\pm 0.25$ m/s. Next, within each interval, the $p(s,\theta)$s  were sorted by ascending values of $s$, irrespective of the value of $\theta$, and binned in ranges $[s_i-\Delta_i s/2,s_i+\Delta_i s/2]$ containing a fixed number of points (typically, a few thousands), hence using a variable width $\Delta_i s$. Then, for each $s_i$,  $a_0(s_i)$, $a_1(s_i)$ and $a_2(s_i)$ were obtained from a least-square fit of the measured data using Eq. (\ref{azimfit}), leading to the typical results (red lines) shown in Fig. \ref{fig:exemplefit}. Finally, the original IASI pdf $p(s_u,s_c)$ for each wind speed was corrected and replaced by $\tilde p(s_u,s_c)=\max(p(s_u,s_c) -b,0)$, with $b$ such that:
 \be\label{normpdf}
 \int_{-\infty}^{+\infty}\int_{-\infty}^{+\infty} \tilde p(s_u,s_c)ds_uds_c = \int_0^\infty 2\pi s\ \tilde a_0(s)ds=1 ,
 \ee
 where $\tilde a_0(s)=max(a_0(s)-b,0)$. This leads to the black lines in the example of Fig. \ref{fig:pdf7up}. As seen, the subtraction of the small estimated bias restores the expected rapidly decreasing tail of the distribution with very little impact on the central part. \textcolor{black}{The retrieved values of $b$ vary between 0.15 and 0.195 and are found wind-dependent, as expected from the wind-dependent nature of the observed bias.} In order to evaluate the slope $s_0$ below which the pdf can be considered as accurate, we used the (classical) criterion that the signal should be at least twice larger than the noise threshold (3 dB SNR), that is $a_0(s)=2b$. This leads to values from $s_0$=0.32 (18$^{\circ}$) at 3 m/s to $s_0$=0.52 (27.5$^{\circ}$) at 15 m/s, which are slightly larger than the maximal ones obtained by \cite{Cox54,Cox54b,Cox56} which range from about 0.27 (15$^{\circ}$) to 0.47 (25$^{\circ}$) depending on wind speed.  Note, however, that the range of available slopes probed by IASI extends far beyond these values, with a significant number of wave tilts exceeding $1$ ($45^\circ$), so that steeper slopes in the pdf can be used after correction of the bias. Nevertheless, our statistical analysis will be mainly based on the most reliable range of slopes, which is limited by the aforementioned values.
 
Finally note that, for the analysis and results presented in Secs. 5.2 and 6.1, respectively, the raw IASI data were replaced (but still denoted as IASI data), for each wind speed, by those for the up-, down-, and cross-wind directions computed for each binned slope $s_i$ by using Eqs. (\ref{pup}) and the values of $a_0(s_i)=\tilde a_0(s_i)$, $a_1(s_i)$, and $a_2(s_i)$ determined as described above. This corrects for the bias in the tail of the distribution and filters the scatter, as shown by comparing the black line with the blue dots in Fig. \ref{fig:pdf7up}.
   
\section{Analysis of the probabilities and modeling the pdf}
We below first recall the main elements of the study of \cite{Cox54,Cox54b,Cox56}. This enables to introduce the main statistical parameters and to discuss some challenging aspects of their estimation. We then present our analysis approach which uses several original techniques.

\subsection{Recalling the Cox and Munk (CM) analysis}
In their celebrated investigations \citep{Cox54,Cox54b,Cox56}, CM derived the wave-slope distribution from the analysis of sun-glitter aerial photographs. To date, their relationships between the up- and cross-wind mean squared slopes  (MSSs) and wind speed still remain the reference and the most frequently used.

Recall that the up- and cross-wind MSSs, $m_u$ and $m_c$, are, for a given wind speed, defined as the second moments of the pdf, i.e.:
\begin{equation}
m_u=\int_{-\infty}^{+\infty} s_u^2p(s_u,s_c)ds_uds_c,\ m_c=\int_{-\infty}^{+\infty} s_c^2p(s_u,s_c)ds_uds_c ,
\label{defsigma}
\end{equation}
the total MSS being $m_t=m_u+m_c$. The main findings of CM concerning $p(s_u,s_c)$ are that: a) It is close to a bidimensional normal distribution, with a null mean slope and $m_u>m_c$ ; b) It is symmetrical with respect to the crosswind direction and negatively skewed with respect to the upwind one ; c) With respect to a Gaussian distribution, it is slightly more peaky around the origin ($s_u=s_c=0$), and decreases more slowly at large slopes. d) $m_u$, $m_c$ and $m_t$ increase quasi-linearly with wind speed. CM also proposed to describe the  deviations from Gaussian by an expansion in terms of a Gram-Charlier (GC) series, i.e.:
\begin{equation}
\begin{split}
&  p(s_u,s_c)=(2\pi\sqrt{m_um_c})^{-1}e^{-\demi u^2}e^{-\demi c^2}\times \left(1-\frac{C_{12}u(c^2-1)}{2}\right.\\
&-\frac{C_{30}(u^3-3u)}{6} +\frac{C_{40}(u^4-6u^2+3)}{24}\\
&\left.+\frac{C_{22}(u^2-1)(c^2-1)}{4}+\frac{C_{04}(c^4-6c^2+3)}{24} \right) , \end{split}
\label{GC}
\end{equation}
where $u=s_u/\sqrt{m_u}$ and $c=s_c/\sqrt{m_c}$. $C_{12}$ and $C_{30}$ are the skewness coefficients along the wind direction, which account for the asymmetry of the distribution, while the kurtosis coefficients $C_{40}$,  $C_{22}$ and $C_{04}$ enable to describe the peakedness. This non-Gaussian attribute has been shown important for the accurate characterization of the near-nadir microwave radar cross-section \citep{chapron_JGR00,Hauser,bringer_GRS12,boisot2015,Chen}.
Note that, for consistence with many studies in radar and optical remote sensing, we chose to define the x-axis, which is the reference of azimuth, as the upwind direction. \textit{We therefore use switched notations with respect to those of CM (and  BH)}, who chose the y axis to point upwind. This has no consequence in the analysis but this implies that the  index $i$ in our $C_{ij}$ refers to the upwind axis while $j$ refers to the crosswind one (hence our $C_{ji}$ corresponds to the $C_{ij}$ of CM and BH).

Since the CM data were limited to small and moderate wave tilts, only ``incomplete'' MSSs could be evaluated due to truncated integrations in Eq. (\ref{defsigma}), limited to some wind-dependent threshold value $s_{0}$. For more accurate estimations, CM carried an extrapolation based on the fit, to their measured data, of an analytical representation of the logarithm of the probabilities using truncated Fourier series of the azimuthal angle $\theta$ (with respect to the upwind direction) and power series of the wave slope $s=(s_u^2+s_c^2)^{1/2}$, i.e.:
\be
\begin{split}
&\log p(s,\theta)=\alpha_0-\alpha_0's^2+\alpha_0''s^4+s(\alpha_1+\alpha_1's^2)\cos\theta\\
&+s^2(\alpha_2+\alpha_2's^2)\cos(2\theta),
\end{split}
\label{logfit}
\ee
A key parameter in this expansion is the coefficient $\alpha_0'$, corresponding to the quadratic term of the omnidirectional (i.e, azimuthally averaged) pdf, which is an inverse MSS ($\alpha_0'= 1/m_t$ under an isotropic Gaussian assumption). This coefficient, which can be estimated quite easily from a fit of the measured data around  the origin ($s=\theta=0$), drives the decay of the main quasi-Gaussian lobe of the distribution and therefore also controls the available range of values for the wave tilts which are limited by the dynamical range of the pdf values above the noise threshold.
with CM experimental data extending to  $s_0\simeq s_4=(4/\alpha_0')^{1/2}$. Assuming Eq. (\ref{logfit}) to hold beyond the limiting slope $s_4$ and determining the coefficients $\alpha_j^{,',''}$ by fitting the data in azimuth ($\theta$) and wave slope ($s$) using Eq. (\ref{logfit}), CM were then able to extrapolate and obtain more reliable MSSs. They observed the convergence of the integrals in  Eq. (\ref{defsigma}) for $s_8=(8/\alpha_0')^{1/2}$ and found wind-independent factors with respect to the incomplete values calculated at $s_4$, i.e.:
\be
\begin{split}
&m_u\simeq 1.23\int_{\abs s\leq s_4} s_u^2p(s_u,s_c)ds_uds_c\\
&m_c\simeq 1.22\int_{\abs s\leq s_4} s_c^2p(s_u,s_c)ds_uds_c .
\end{split}
\label{renorm}
\ee
Finally, the following famous formula were proposed  for wind speeds at the ``mast heigth'`` of 12.5 m in the 1-14 m/s range \citep{Cox54}:
\begin{equation}
\begin{split}
  m_u=3.16\ 10^{-3}\ U_{12.5}\quad,\quad 
  m_c=3.\ 10^{-3} +1.92\ 10^{-3} U_{12.5} .
\end{split}
\label{loiCM}
\end{equation}
The skewness ($C_{12}$, $C_{30}$) and kurtosis ($C_{40}$, $C_{04}$, $C_{22}$)  coefficients were deduced from the fitted coefficients in Eq. (\ref{logfit}) and the MSSs, but with a poor accuracy. Recall that the CM results were confirmed by \cite{Breon} using wave-slope probabilities, retrieved from satellite observations, directly fitted with the GC model of Eq. (\ref{GC}) floating 7 parameters (the 2 directional MSSs, the 2 skewness and 3 kurtosis coefficients). The values of $m_u$ and $m_c$  obtained are very close to those of CM while some GC coefficients show a dependence on wind speed that CM could not establish due to insufficient accuracy. Our attempts to carry similar fits of the IASI-retrieved probabilities led to ambiguous solutions due to the numerous floated parameters and the significant correlations between them. We found, in particular, that variations in the estimated MSSs and kurtosis parameters can compensate each other through the quadratic terms in Eq. (\ref{GC}) and can only be separated by the quartic terms that enter in play for larger values of the wave tilt, so that the optimization procedure is not sufficiently robust and can lead to an ambiguous set of solutions. In order to overcome this problem, We devised an adapted and original procedure for the evaluation of the MSSs and GC coefficients, which is described below.

\subsection{An original  approach for the analysis of the slope probabilities}\label{section:procedure}

Retaining the GC expansion in Eq. (\ref{GC}) we here present a methodology for robust estimations of the directional MSSs as well as of the skewness and kurtosis coefficients. Our approach relies on the use of the azimuthal expansion [Eq. (\ref{azimfit})] to fit the measured distribution  rather than its logarithm, as was done by CM [Eq. (\ref{logfit}]. This combines several advantages: a) Taking the logarithm of the distribution can produce very large negative values that give an excessive weight to small probabilities in a least-square fitting of the Fourier harmonics; b) The moments of the azimuthal functions are useful for the estimation of the directional MSSs and  GC coefficients, as shown later; c) This representation enables to separate the symmetric and anti-symmetric components of the alongwind distribution:
    \begin{equation}
      \label{psym}
      \begin{split}
        p_{sym}(s_u,s_c)&=\frac{p(s_u,s_c)+p(-s_u,s_c)}{2}=a_0(s)+a_2(s)\cos2\theta,\\
        p_{asym}(s_u,s_c)&=\frac{p(s_u,s_c)-p(-s_u,s_c)}{2}=a_1(s)\cos\theta,
         \end{split}
    \end{equation}
    which facilitates the identification of the skewness and kurtosis coefficients in the GC model as $p_{sym}$ (resp. $p_{asym}$) is obtained by setting $C_{12}=C_{30}=0$ (resp. $C_{40}=C_{04}=C_{22}=0$) in Eq. (\ref{GC}).
  
\subsubsection{Estimation of the mean square slopes (MSSs)}
As discussed above, the estimation of the MSSs from experimentally-determined wave-slope probabilities is a challenging problem due to the uncertainties affecting the measured pdf and the lack of reliable data beyond a given slope, which led CM to use extrapolations and a renormalization, a procedure which has been criticized by further analysis \citep{wentz_JGR76}. We propose a different and original approach based on a direct calculation (and not a fit) of the MSSs using an appropriate combination of moments $\cali M_{kn}$ of the azimuthal functions, defined by:
\be
{\cali M}_{kn}=\int_0^\infty s^k\ a_n(s)ds .
\label{defmoments}\ee
Note that the normalization  of the pdf implies:
\be\label{normcond}
2\pi {\cali M}_{10}=1,
\ee
and that the directional MSSs are:
\be
\begin{split}
m_u=\pi ({\cali M}_{30}+{\cali M}_{32}/2)\quad,\quad
m_c=\pi ({\cali M}_{30}-{\cali M}_{32}/2).
\end{split}
\label{mssdirect}
\ee
As discussed in Sec. 4, the IASI-retrieved probabilities are potentially biased (and significantly contaminated by noise) for wave-tilt angles greater than 18$^{\circ}$-34$^{\circ}$ depending on the wind speed. For these limit values, the integrals defining the moments [Eq. (\ref{defmoments})] are not always converged. Furthermore, since the relative contribution of steep  waves to  ${\cali M}_{kn}$ increases  with increasing $k$, so does the bias on ${\cali M}_{kn}$ due to truncation, and using the third-order ones ${\cali M}_{30}$ and ${\cali M}_{32}$ to determine the MSSs from Eq. (\ref{mssdirect}) is thus inappropriate. This led us to consider some lower moments of the distribution also involving the MSSs. Indeed, one has:
\be
\begin{split}
  \int_{0}^{+\infty}  p_{sym}(s_u,0)ds_u&={\cali M}_{00}+{\cali M}_{02}=\demi(2\pi m_c)^{-1/2}\left(1+\frac{C_{04}}{8}\right),\\
  \int_{0}^{+\infty}  s_u p_{sym}(s_u,0)ds_u&={\cali M}_{10}+{\cali M}_{12}= \sqrt{\frac{m_u}{2\pi}}(2\pi m_c)^{-1/2}\\
  &\times \left(1+\frac{1}{8}C_{04}-\frac{1}{24}C_{40}-\frac{1}{4}C_{22}\right),\\
\int_{0}^{+\infty}  p_{sym}(0,s_c)ds_c&={\cali M}_{00}-{\cali M}_{02}=\demi(2\pi m_u)^{-1/2}\left(1+\frac{C_{40}}{8}\right),\\
\int_{0}^{+\infty}  s_c p_{sym}(0,s_c)ds_c&={\cali M}_{10}-{\cali M}_{12}= \sqrt{\frac{m_c}{2\pi}}(2\pi m_u)^{-1/2}\\
&\times \left(1+\frac{1}{8}C_{40}-\frac{1}{24}C_{04}-\frac{1}{4}C_{22}\right),\\
\end{split}
\ee
from which one obtains:
\be\begin{split}
m_u=\frac{\pi}{2}F_u\times \left(\frac{{\cali M}_{10}+{\cali M}_{12}}{{\cali M}_{00}+{\cali M}_{02}}\right)^2 \quad,\quad
m_c=\frac{\pi}{2}F_c\times \left(\frac{{\cali M}_{10}-{\cali M}_{12}}{{\cali M}_{00}-{\cali M}_{02}}\right)^2,
\end{split}
\label{calculemssfin}\ee
with
\be
\begin{split}
  &F_u=\left(\frac{1+\frac{1}{8}C_{04}}{1+\frac{1}{8}C_{04}-\frac{1}{24}C_{40}-\frac{1}{4}C_{22}}\right)^2,\\
  &F_c=\left(\frac{1+\frac{1}{8}C_{40}}{1+\frac{1}{8}C_{40}-\frac{1}{24}C_{04}-\frac{1}{4}C_{22}}\right)^2.
  \end{split}
\label{Ffactors}\ee
Anticipating that $F_u$ and $F_c$ have values very close to $1$ (as confirmed later) we obtain, using Eqs. (\ref{normcond}) and (\ref{calculemssfin}): 
\be
\begin{split}
  m_u\simeq m_u^0= \frac{1}{8\pi}\left(\frac{1+2\pi {\cali M}_{12}}{{\cali M}_{00}+{\cali M}_{02}}\right)^2 \quad,\quad
  m_c\simeq m_c^0= \frac{1}{8\pi}\left(\frac{1-2\pi {\cali M}_{12}}{{\cali M}_{00}-{\cali M}_{02}}\right)^2.
\end{split}
\label{calculemss}
\ee
 These equations involve moments of order $k=0$ and $1$, which converge rapidly and thus provide  good first guesses ($m_u^0$ and $m_c^0$) of $m_u$ and $m_c$ (which will be refined after estimation of the factors $F_u$ and $F_c$). A careful investigation of the cumulative sums involved in the  moments shows that $m_u^0$ and $m_c^0$ in Eq. (\ref{calculemss}) are indeed correctly evaluated at the limiting slope of the IASI data, which is not the case when Eq. (\ref{mssdirect}) is used. 

\subsubsection{Estimation of the kurtosis coefficients}

In order to determine $F_u$ and $F_c$ from Eq. (\ref{Ffactors}), we must estimate $C_{40}$, $C_{04}$, and $C_{22}$ which appear in Eq. (\ref{GC}). Those are actually the excess kurtosis of the two-dimensional distribution, given by:

\be
\label{kurt}
\begin{split}
C_{40}&=  m_u^{-2}\int_{-\infty}^{+\infty}\int_{-\infty}^{+\infty} s_u^4 p(s_u,s_c)ds_uds_c-3,\\
C_{04}&=  m_c^{-2}\int_{-\infty}^{+\infty}\int_{-\infty}^{+\infty} s_c^4p(s_u,s_c)ds_uds_c-3,\\
C_{22}&=  (m_um_c)^{-1}\int_{-\infty}^{+\infty}\int_{-\infty}^{+\infty} s_u^2s_c^2p(s_u,s_c)ds_uds_c-1,
\end{split}
\ee

which involve moments, of orders 4 and 5 in Cartesian and polar coordinates, respectively, which cannot be reliably calculated  due to the uncertainties affecting the tail of the distribution. Nevertheless, $C_{40}$, $C_{04}$, and $C_{22}$ can also be obtained from direct fits of the IASI probabilities using the GC model, provided that the MSSs are known and can be fixed. This was done using the values of $m_u^0$ and $m_c^0$ obtained from Eq. (\ref{calculemss}) as first guesses, before refining the estimations. We thus determined $C_{40}$, $C_{04}$, and $C_{22}$ through least-square adjustments of the measured symmetrical components $p_{sym}$ along the 3 principal directions ($\theta=0^\circ,\ 90^\circ,\ 45^{\circ}$) using Eq. (\ref{psym}) and the associated expressions of the GC model. Then,  the final MSSs were obtained from:
\be\label{mymss}
\begin{split}
  m_u=F_u(m_u^0,m_c^0)\times m_u^0 \qquad , \qquad
  m_c=F_c(m_u^0,m_c^0)\times m_c^0,
  \end{split}
\ee
based on factors $F_u$ and $F_c$ computed using Eq. (\ref{Ffactors}) and the obtained values of $C_{40}(m_u^0,m_c^0)$, $C_{04}(m_u^0,m_c^0)$ and $C_{22}(m_u^0,m_c^0)$. Finally, refitting the kurtosis coefficients on the basis of the refined MSSs yielded the definitive best estimates $C_{40}(m_u,m_c)$, $C_{04}(m_u,m_c)$ and $C_{22}(m_u,m_c)$. Based on our estimated kurtosis coefficients (see further), we found the corrective factors $F_u$ and $F_c$ very close to $1$ (within $3\%$) so that the approximated MSSs of Eq. (\ref{calculemss}) are very close to the refined ones in Eq. (\ref{calculemssfin}).
 
\subsubsection{Estimation of the skewness coefficients}

The skewness of the distribution is accounted for by the $a_1$ function in Eq. (\ref{azimfit}).  Similarly to the kurtosis coefficients, the skewness parameters $C_{30}$ and $C_{12}$ can be obtained by fitting the asymmetric component $p_{asym}$ of the measured distribution derived from Eq. (\ref{psym}) with that of the GC model once the MSSs are known. However, they can also be derived from a direct calculation using the moment ${\cali M}_{11}$, since one has:
\be
\begin{split}\label{C12moment}
  \int_{-\infty}^{+\infty} s_u\ p_{asym}(s_u,0)ds_u&=2M_{11}=C_{12}\frac{1}{\sqrt{2\pi}} \left(\frac{m_u}{m_c}\right)^{1/2}, \\
  \int_{-\infty}^{+\infty}  \mathrm{sign}(s_u)\ p_{asym}(s_u,s_c)ds_uds_c&=4M_{11}=\frac{2}{3}C_{30}\frac{1}{\sqrt{2\pi}},
\end{split}
\ee

\be
  C_{12}=2\sqrt{2\pi}\left(\frac{m_c}{m_u}\right)^{1/2}M_{11} \qquad , \qquad  
  C_{30}=6\sqrt{2\pi}M_{11}.
\ee
Note that this implies that :
\be
C_{30}/C_{12}=3\sigma_u/\sigma_c,
\ee
a relation involving the directional slopes root mean squares $\sigma_{u,c}=m^{1/2}_{u,c}$ which is to our best knowledge original.

\subsubsection{Estimation of the MSS shape}

Useful quantities in optical and microwave remote sensing of the sea surface are the so-called ``MSS-shape'' \citep{jackson1992,chapron_JGR00,SWH1}, obtained by fitting the measured pdf with a Gaussian distribution. They are primary parameters to describe the data, which are more easily determined than $m_u$ and $m_s$ and appear naturally when using the well-known Geometrical Optics model \citep{hagfors1966} in near-nadir scattering. Similarly to the up- and cross-wind MSSs, we define an upwind ($m_{us}$) and a downwind ($m_{cs}$) MSS-shape by:
\begin{equation}
p(s_u,s_c)/p(0,0)\simeq e^{-\demi s_u^2/m_{us}}\ e^{-\demi s_c^2/m_{cs}}.
\label{defmssshape}
\end{equation} 
Expansions of Eqs. (\ref{GC}) and (\ref{defmssshape})  to second-order in slope and identification of the quadratic terms shows that $m_{us}$ and $m_{cs}$ are related to $m_u$ and $m_c$ by:
\be
\begin{split}
  m_{us}&=\left(1+\frac{C_{40}+C_{22}}{2+C_{04}/4+C_{40}/4+C_{22}/2}\right)^{-1}\ m_u ,\\
 m_{cs}&=\left(1+\frac{C_{04}+C_{22}}{2+C_{04}/4+C_{40}/4+C_{22}/2}\right)^{-1}\ m_c .\\
  \end{split}
  \label{mssshapetomss}
\ee
$m_{us}$ and $m_{uc}$ are thus in general smaller than $m_u$ and $m_c$, respectively, with differences resulting for the  non-Gaussian nature of the distribution due to its peakedness (which is found most often with a positive kurtosis). The MSS-shape can be easily and robustly determined since they do not require an absolute calibration of the probabilities and are quite insensitive to the uncertainties affecting the tail of the pdf. Indeed, Eq. (\ref{defmssshape}) implies that the functions $s_u^2p(s_u,0)/p(0,0)$ and  $s_c^2p(0,s_c)/p(0,0)$ reach their maxima, of $2m_{us}e^{-1}$ and $2 m_{uc}e^{-1}$, at $s_u=\sqrt{2 m_{us}}$ and $s_c=\sqrt{2 m_{uc}}$, respectively, which are only at 1.4 standard deviation away from $s_u=s_c=0$.

\section{Results and discussion}
\subsection{Results}

\begin{figure}
  \begin{minipage}[c]{0.5\textwidth}
 \includegraphics[width=1\textwidth]{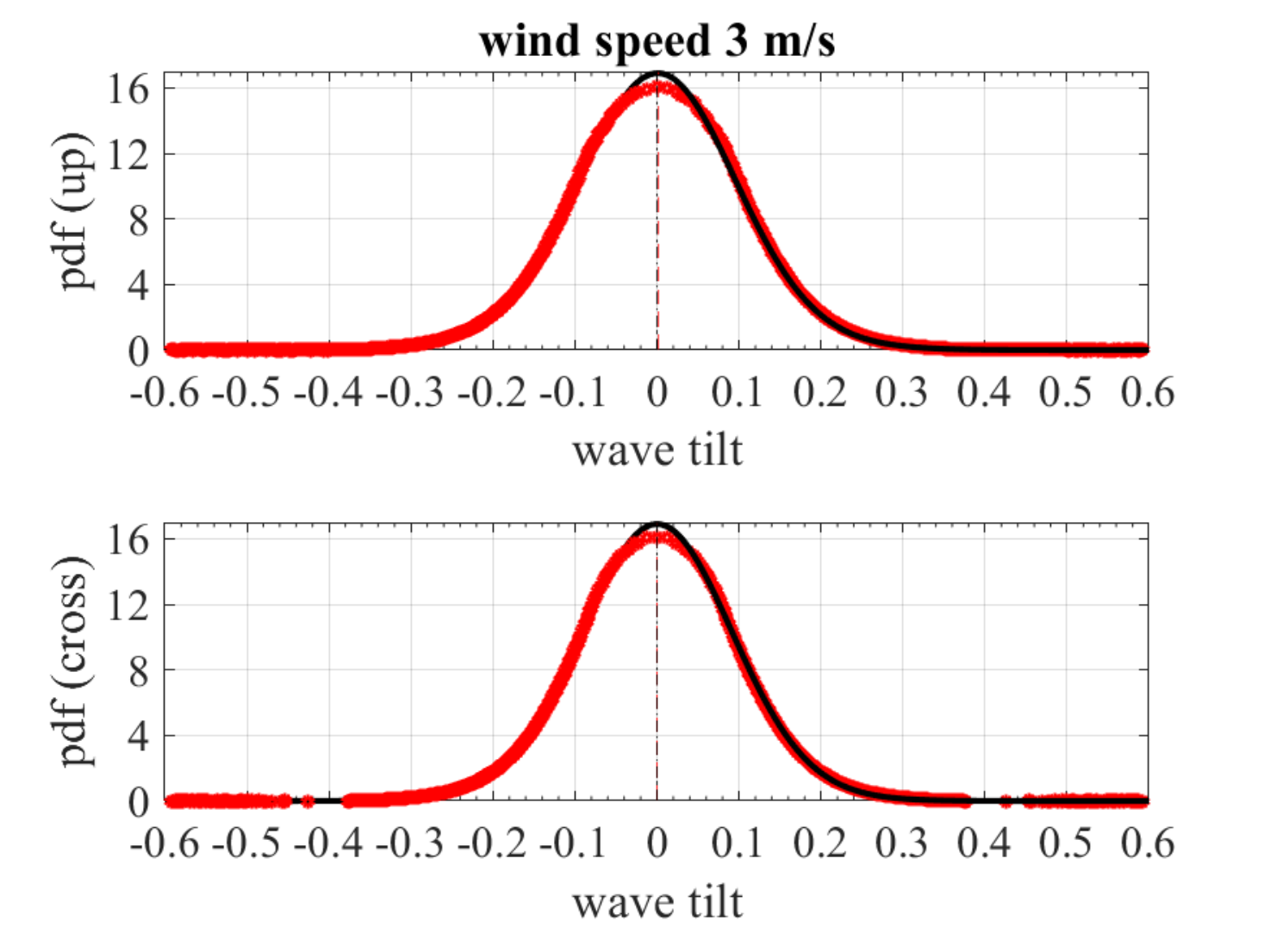}
\end{minipage}\begin{minipage}[c]{0.5\textwidth}
 \includegraphics[width=1\textwidth]{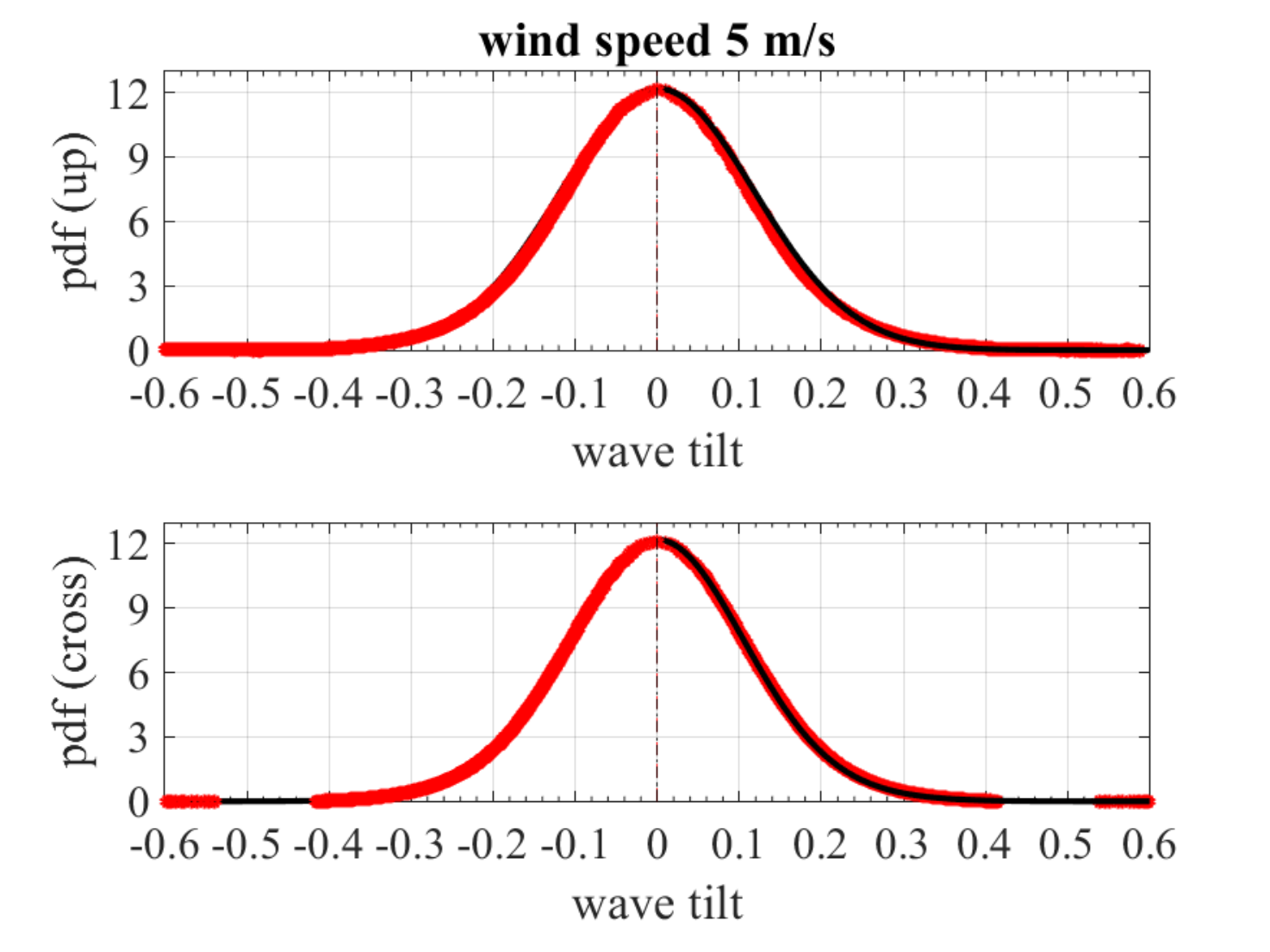}
  \end{minipage}
  
 \begin{minipage}[c]{0.5\textwidth}
 \includegraphics[width=1\textwidth]{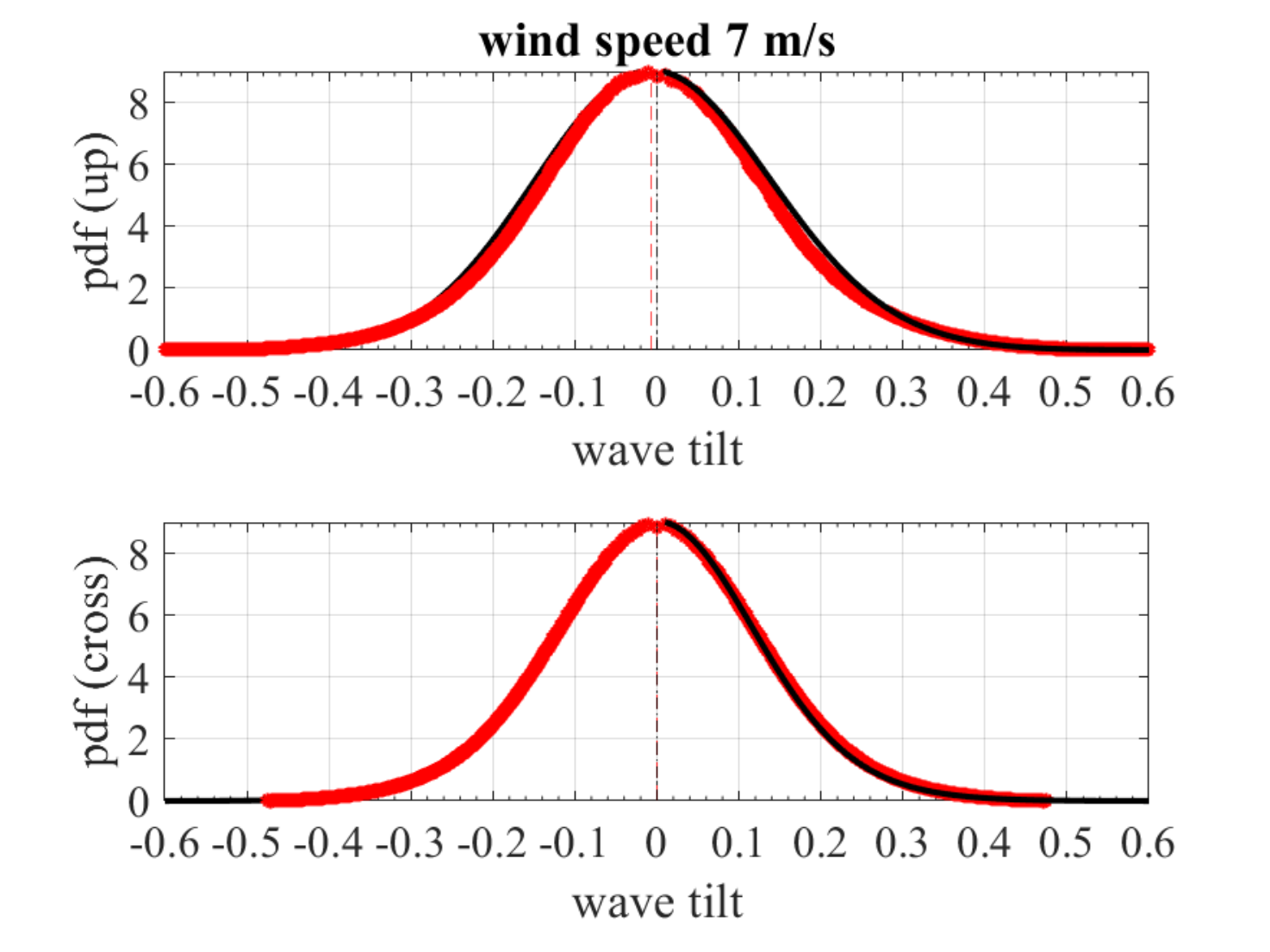}
\end{minipage}\begin{minipage}[c]{0.5\textwidth}
 \includegraphics[width=1\textwidth]{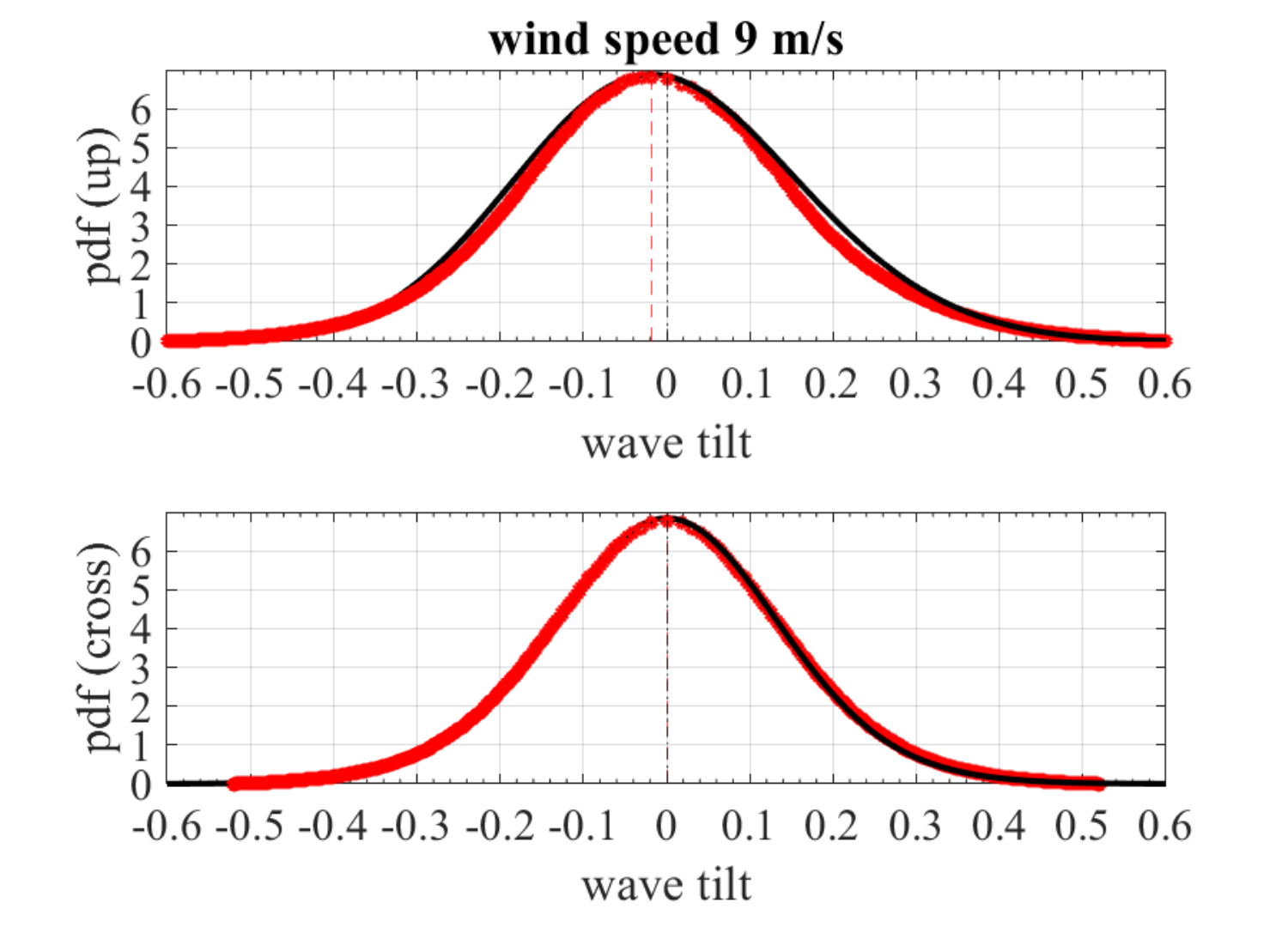}
 \end{minipage}

 \begin{minipage}[c]{0.5\textwidth}
 \includegraphics[width=1\textwidth]{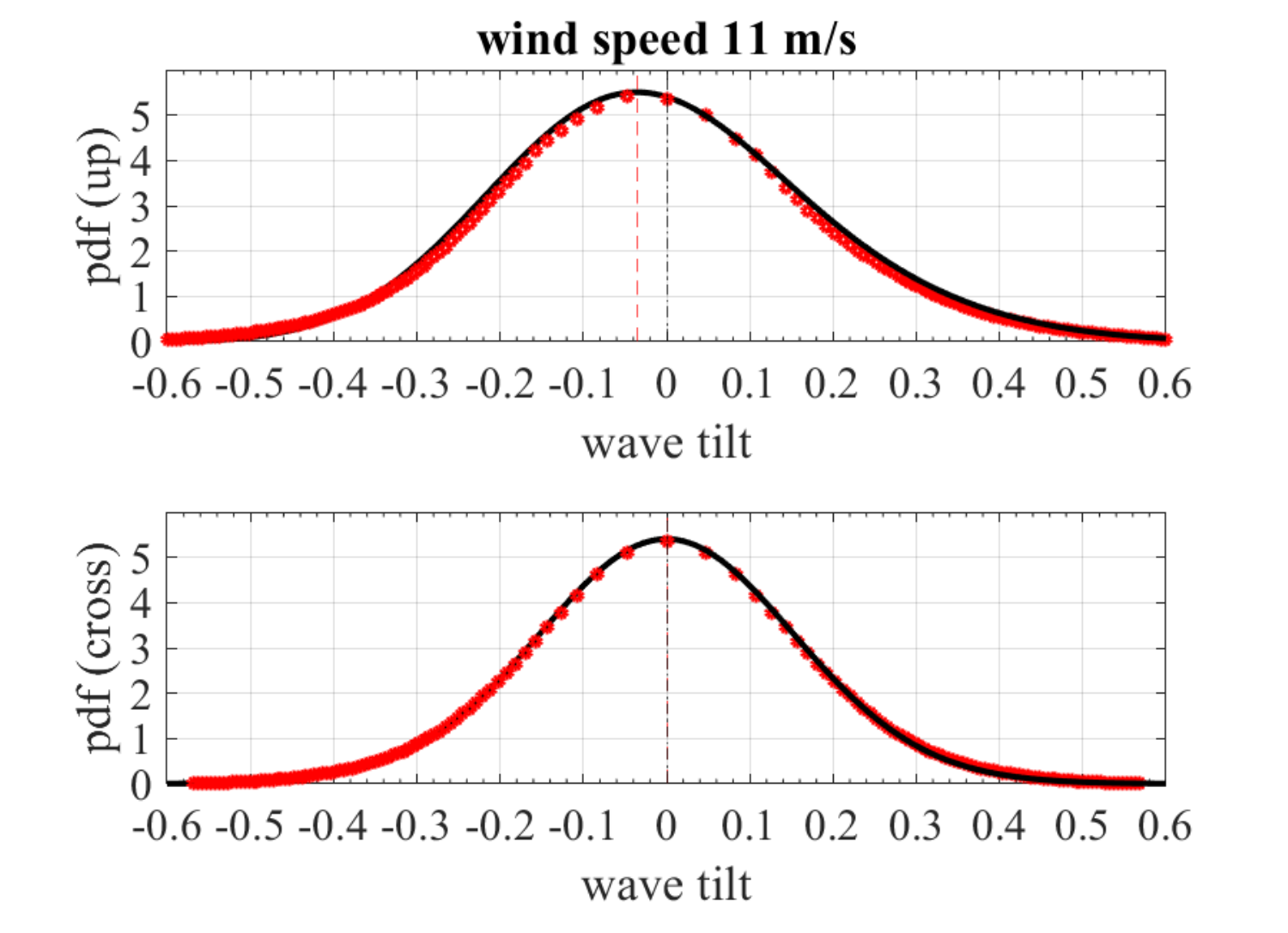}
\end{minipage}\begin{minipage}[c]{0.5\textwidth}
 \includegraphics[width=1\textwidth]{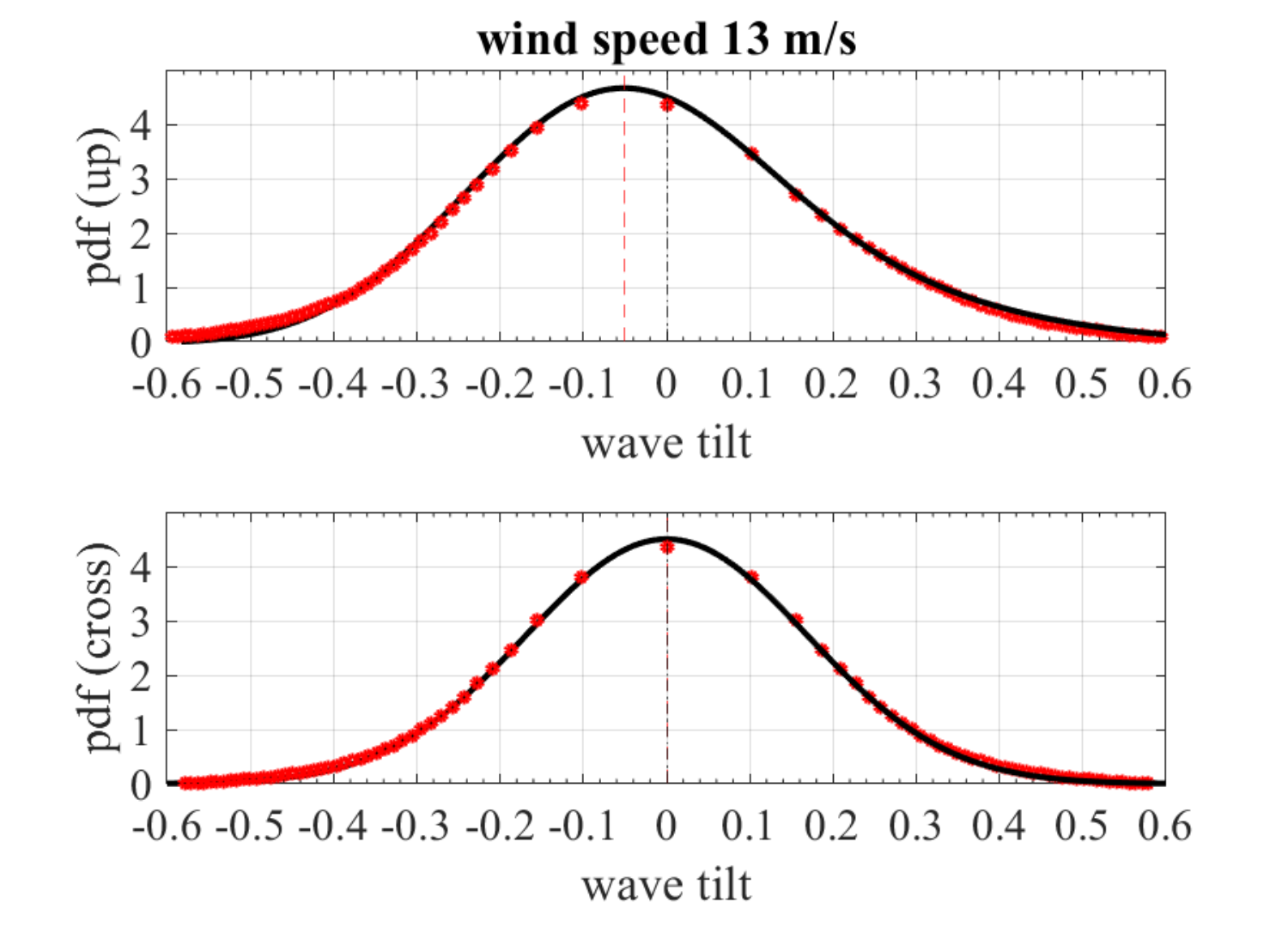}
 \end{minipage}

 \caption{IASI-measured (red dots) along- and cross-wind slope distributions of slopes for different wind speeds and the values (black lines) computed using Eq. (\ref{GC}) with the values in Table \ref{tablemss}.}
   \label{fig:GCfit}
  \end{figure}

The parameters feeding Eq. (\ref{GC}), determined as explained in Sec. 5.2 from the probabilities retrieved from IASI observations, are given in Table \ref{tablemss}. They include the directional MSSs ($m_u$ and $m_c$), the kurtosis coefficients ($C_{40}$, $C_{04}$, and $C_{22}$) and the skewness  ($C_{12}$ and $C_{30}$) coefficients. We also give the directional MSS-shape ($m_{us}$ and $m_{cs}$). Note that no results are given for wind  speeds $U_{10} < 3$ m/s as the observed distributions for slower winds are too far from a Gaussian shape and could not be properly described by a GC model, due to both an increased scatter of the measured probabilities and a reduced number of observations.
As shown by Fig. \ref{fig:GCfit}, the values in Table \ref{tablemss} and Eq. (\ref{GC}) enable a very satisfactory description of the measured probabilities. In particular, the progressive change, with increasing wind speed, of the position of the maximum of the alongwind probabilities is well taken into account. This shift can be quantified by the wave-tilt angle $\theta_{pmax}$ at the position of the maximum, for which we provide wind-speed dependent values (Table \ref{tablemss}). Note that the latter are perfectly consistent with the fit made by \cite{chapron_skewness} (their Figure 1) after the data of CM.

  \begin{table*}\centering
  \begin{tabular}{| l | l | l | l | l | l  | l | l | l | l | l | l |}
    \hline
    $U_{10}$ & $ 100 m_u$ &  $100 m_{c}$ & $100 m_{us}$ & $100 m_{cs}$ & $C_{40}$ & $C_{04}$ & $C_{22}$ & $C_{12}$ & $C_{30}$ & $\theta_{pmax}$ \\
    \hline
3.0 & 1.10 & 0.97 & 1.02 & 0.89 & 0.16 & 0.28 & 0.16 & 0.01 & 0.02 & 0.1 \\
3.5 & 1.16 & 1.03 & 1.06 & 0.97 & 0.18 & 0.18 & 0.08 & 0.01 & 0.02 & 0.1 \\
4.0 & 1.24 & 1.11 & 1.13 & 1.05 & 0.21 & 0.11 & 0.05 & 0.01 & 0.02 & 0.1 \\
4.5 & 1.34 & 1.20 & 1.21 & 1.14 & 0.24 & 0.08 & 0.03 & 0.01 & 0.02 & 0.1 \\
5.0 & 1.46 & 1.28 & 1.30 & 1.22 & 0.26 & 0.08 & 0.02 & 0.00 & 0.01 & 0.0 \\
5.5 & 1.59 & 1.37 & 1.41 & 1.31 & 0.27 & 0.08 & 0.02 & -0.00 & -0.01 & -0.1 \\
6.0 & 1.74 & 1.45 & 1.52 & 1.39 & 0.29 & 0.05 & 0.02 & -0.01 & -0.03 & -0.1 \\
6.5 & 1.91 & 1.53 & 1.67 & 1.47 & 0.32 & 0.03 & 0.01 & -0.02 & -0.05 & -0.2 \\
7.0 & 2.09 & 1.62 & 1.83 & 1.55 & 0.32 & 0.01 & 0.01 & -0.02 & -0.07 & -0.4 \\
7.5 & 2.28 & 1.70 & 1.99 & 1.64 & 0.31 & 0.01 & 0.00 & -0.03 & -0.10 & -0.5 \\
8.0 & 2.49 & 1.79 & 2.18 & 1.74 & 0.32 & -0.01 & -0.01 & -0.03 & -0.12 & -0.7 \\
8.5 & 2.68 & 1.88 & 2.35 & 1.84 & 0.30 & -0.03 & -0.02 & -0.04 & -0.14 & -0.9 \\
9.0 & 2.86 & 1.98 & 2.53 & 1.93 & 0.27 & -0.03 & -0.03 & -0.05 & -0.17 & -1.0 \\
9.5 & 3.05 & 2.08 & 2.72 & 2.04 & 0.24 & -0.04 & -0.04 & -0.06 & -0.20 & -1.3 \\
10.0 & 3.23 & 2.20 & 2.92 & 2.16 & 0.21 & -0.02 & -0.03 & -0.06 & -0.24 & -1.5 \\
10.5 & 3.42 & 2.33 & 3.14 & 2.27 & 0.14 & -0.01 & -0.02 & -0.07 & -0.27 & -1.7 \\
11.0 & 3.60 & 2.46 & 3.35 & 2.42 & 0.09 & -0.00 & -0.00 & -0.09 & -0.32 & -2.1 \\
11.5 & 3.79 & 2.59 & 3.59 & 2.56 & 0.05 & -0.01 & 0.02 & -0.10 & -0.36 & -2.3 \\
12.0 & 3.98 & 2.72 & 3.84 & 2.71 & 0.00 & 0.02 & 0.04 & -0.11 & -0.39 & -2.5 \\
12.5 & 4.13 & 2.83 & 4.04 & 2.83 & -0.03 & -0.01 & 0.05 & -0.12 & -0.43 & -2.8 \\
13.0 & 4.40 & 3.00 & 4.27 & 3.03 & -0.01 & 0.02 & 0.11 & -0.13 & -0.46 & -2.9 \\
13.5 & 4.46 & 3.06 & 4.40 & 3.12 & -0.04 & -0.00 & 0.08 & -0.13 & -0.45 & -3.0 \\
14.0 & 4.52 & 3.10 & 4.30 & 3.07 & -0.01 & -0.09 & 0.04 & -0.12 & -0.43 & -3.0 \\
14.5 & 4.75 & 3.26 & 4.67 & 3.32 & -0.05 & -0.04 & 0.10 & -0.12 & -0.44 & -3.0 \\
15.0 & 4.86 & 3.36 & 4.81 & 3.45 & -0.01 & -0.03 & 0.10 & -0.11 & -0.41 & -2.9 \\

\hline
  \end{tabular}
  \caption{Values of the up- ($m_u$) and cross-wind ($m_c$) MSSs and the GC coefficients versus the wind speed $U_{10}$(m/s) at 10 m above the surface. Also given are the MSS-shape ($m_{us},m_{cs}$), and the wave-tilt angle ($\theta_{pmax}$, in degree) at the maximum probability. The MSSs are given in percent (i.e. multiplied by $100$).}
  \label{tablemss}
  \end{table*}
  
\begin{figure}
  \begin{minipage}[c]{0.5\textwidth}
 \includegraphics[width=1\textwidth]{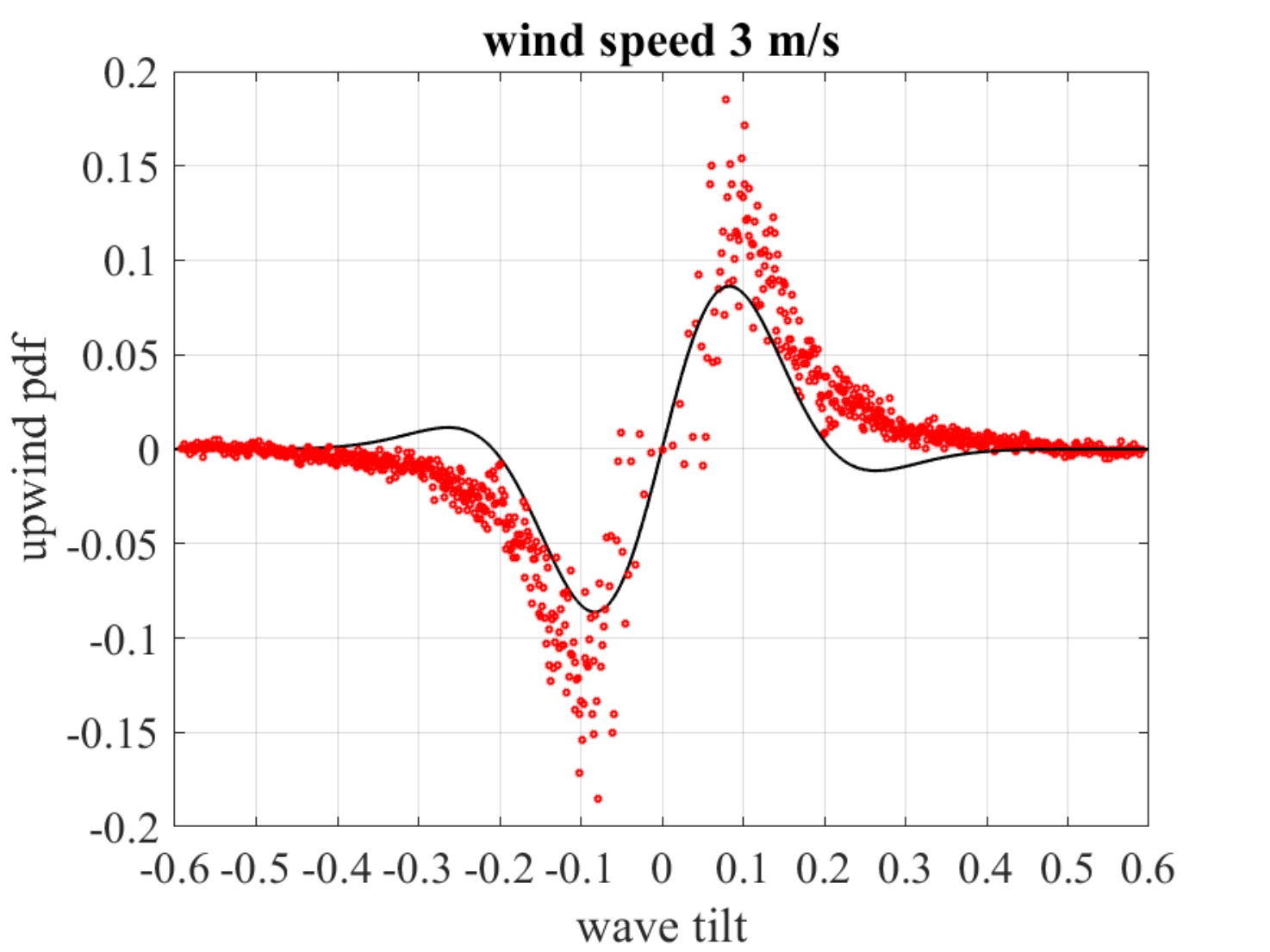}
\end{minipage}\begin{minipage}[c]{0.5\textwidth}
 \includegraphics[width=1\textwidth]{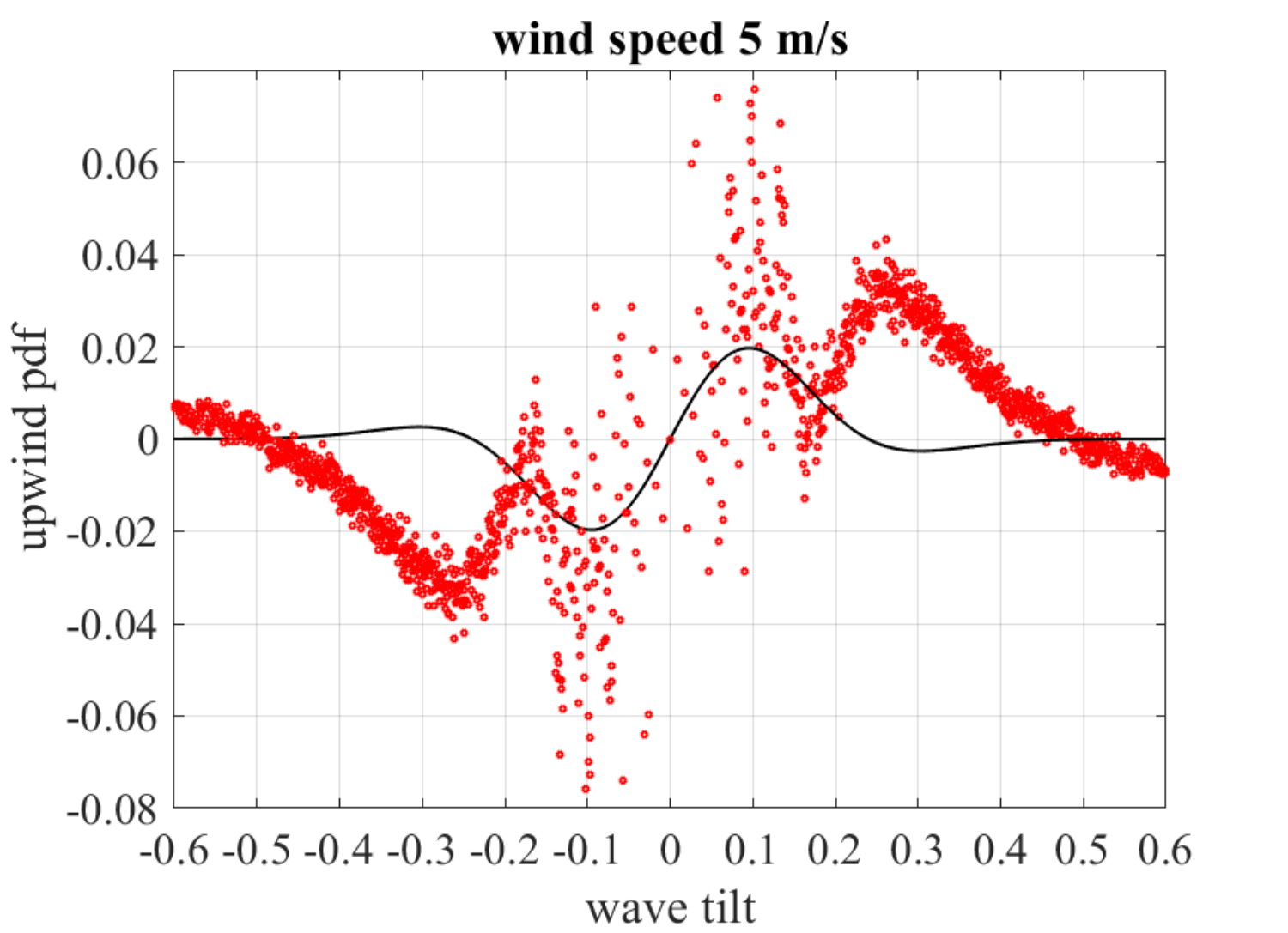}
  \end{minipage}
  
 \begin{minipage}[c]{0.5\textwidth}
 \includegraphics[width=1\textwidth]{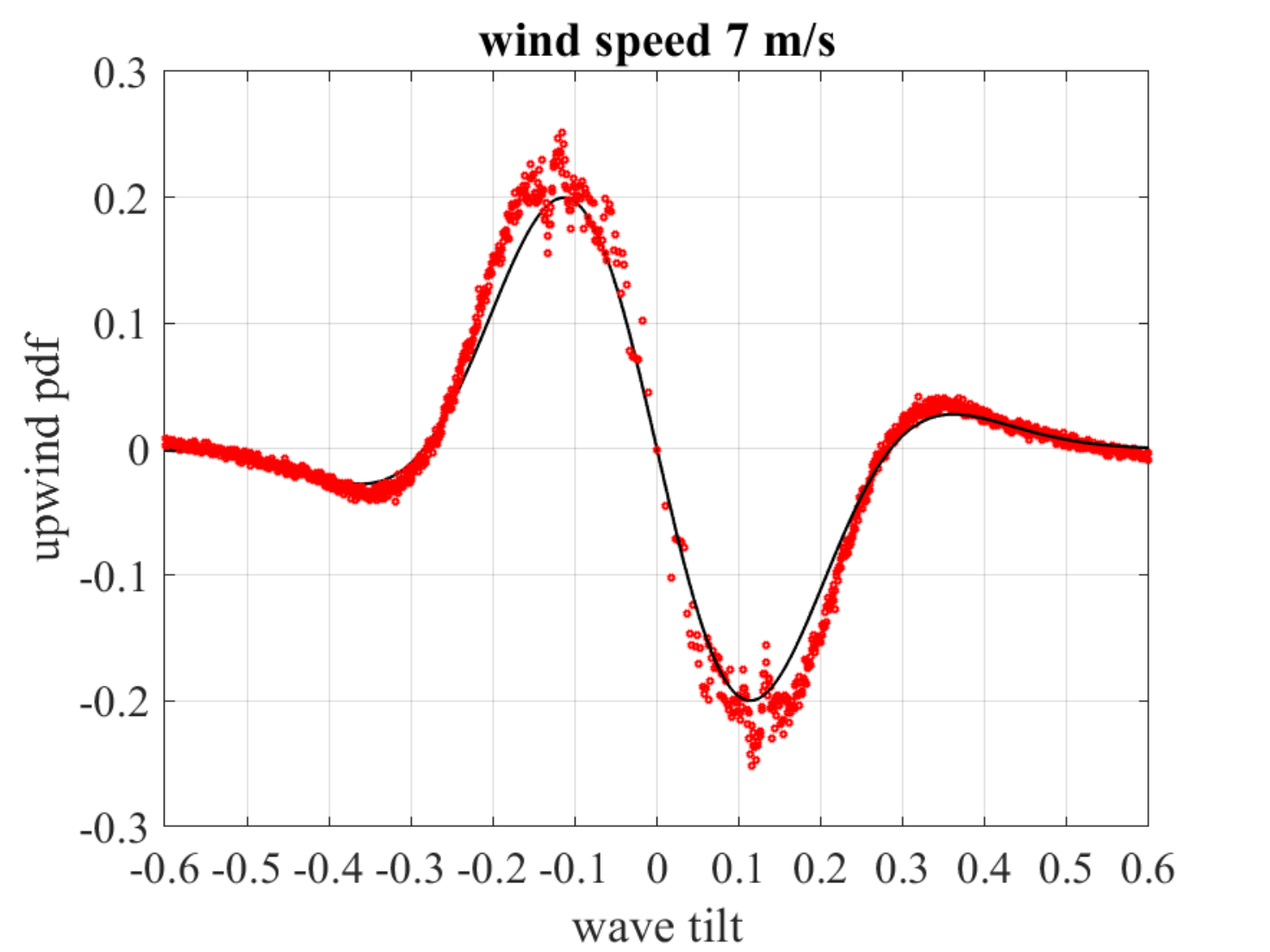}
\end{minipage}\begin{minipage}[c]{0.5\textwidth}
 \includegraphics[width=1\textwidth]{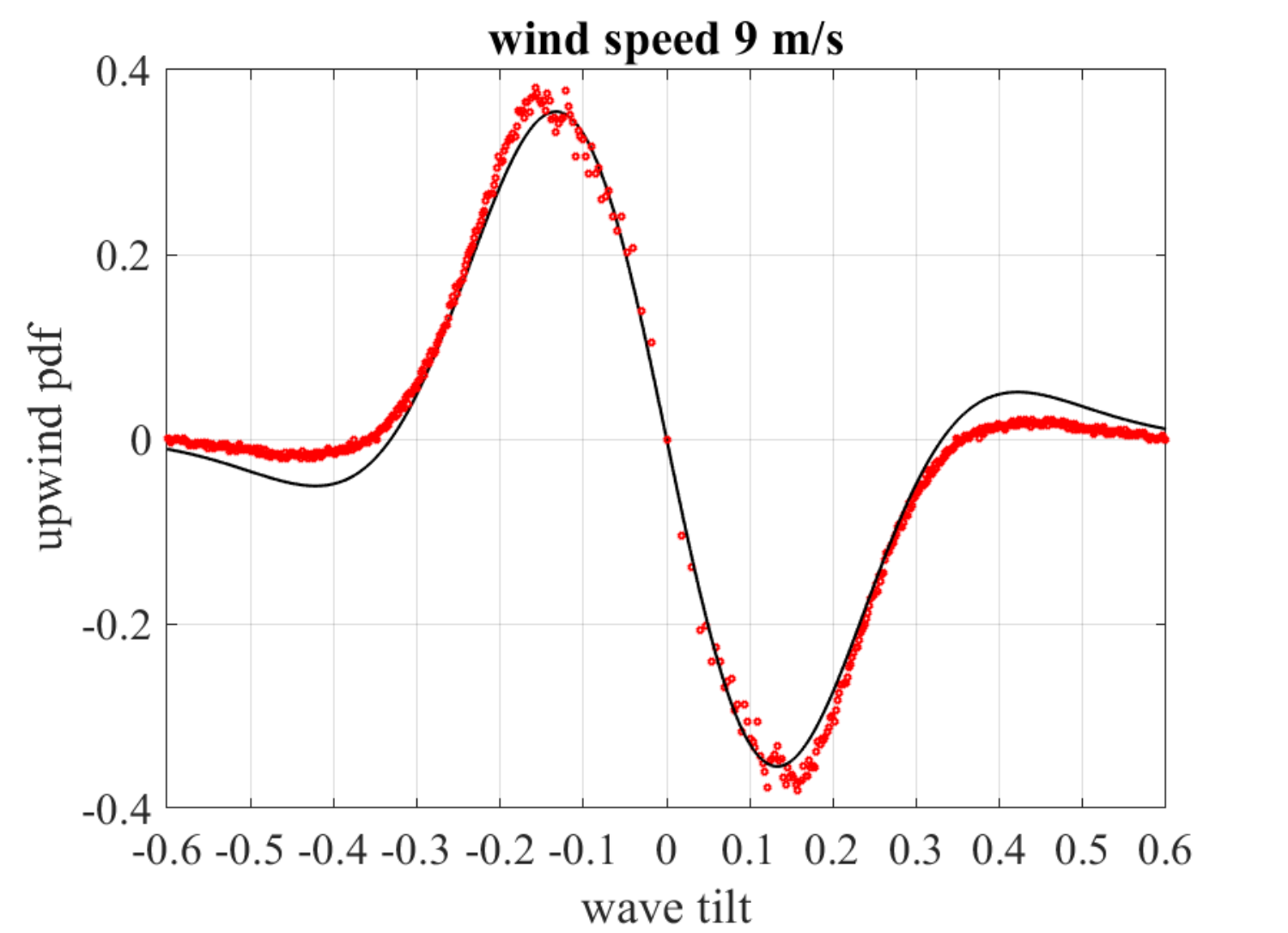}
 \end{minipage}

 \begin{minipage}[c]{0.5\textwidth}
 \includegraphics[width=1\textwidth]{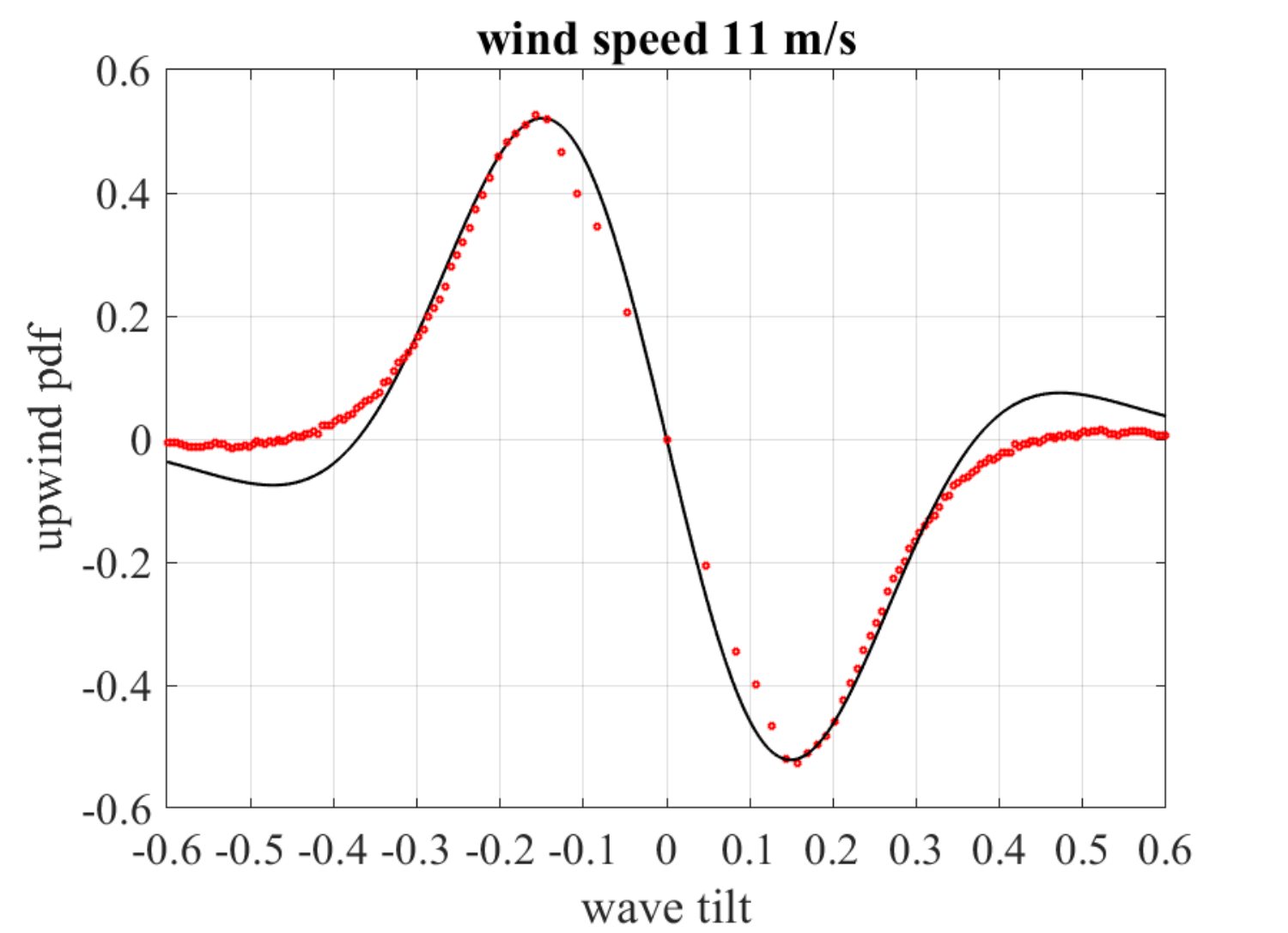}
\end{minipage}\begin{minipage}[c]{0.5\textwidth}
 \includegraphics[width=1\textwidth]{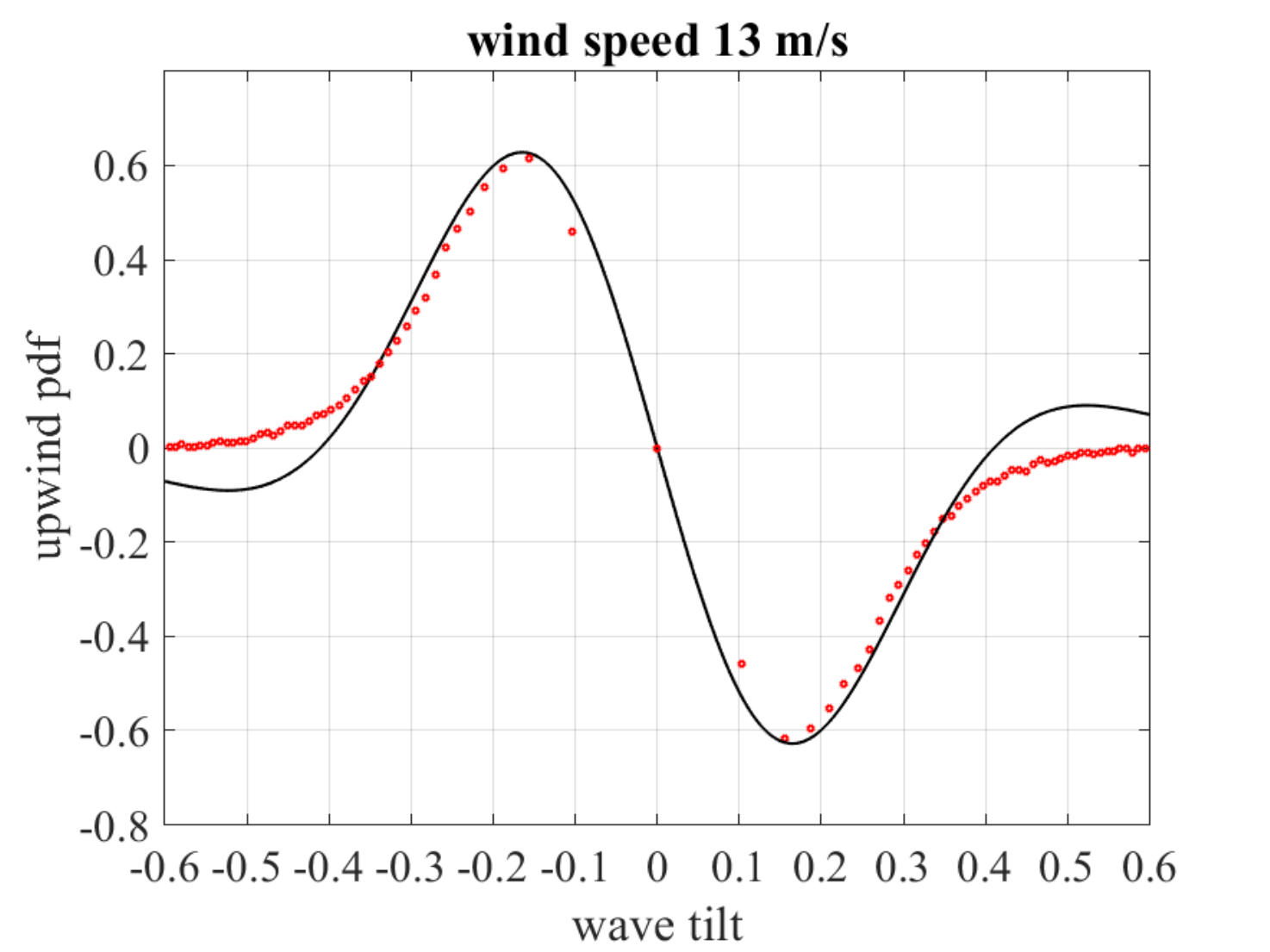}
 \end{minipage}

 \caption{Asymmetric component $p_{asym}$, in the principal section $s_c=0$, determined, using Eq.  (\ref{psym}), from the IASI data (red dots) and values (black lines) computed using Eq. (\ref{GC}) with the values in Table \ref{tablemss}.}
   \label{fig:GCfitas}
  \end{figure}

A more detail look at the asymmetry can be taken by considering the values of $p_{asym}(s_u,s_c)$ [see Eq. (\ref{psym})] in the principal section $s_c=0$ displayed in Fig. \ref{fig:GCfitas}, which call for two remarks. The first is that our model for the pdf again leads to a very satisfactory agreement with the IASI-derived values for this relatively small effect of the wind on the wave slopes. The second is that there is a qualitative evolution when wind speed increases, with a change of sign and a switch from a positive oscillation (below 5 m/s) to a negative one (above 6 m/s). An important consequence of this is that the upwind pdf is larger (resp. smaller) than the downwind one at smaller (resp. larger) wind speeds. We have no explanation for this change, pointed out here for the first time, which indicates a qualitative change in the asymmetry of wave facets as they evolve under complex hydrodynamical processes and wind-surface interactions.

\begin{figure}
   \centering
 \includegraphics[scale=0.25]{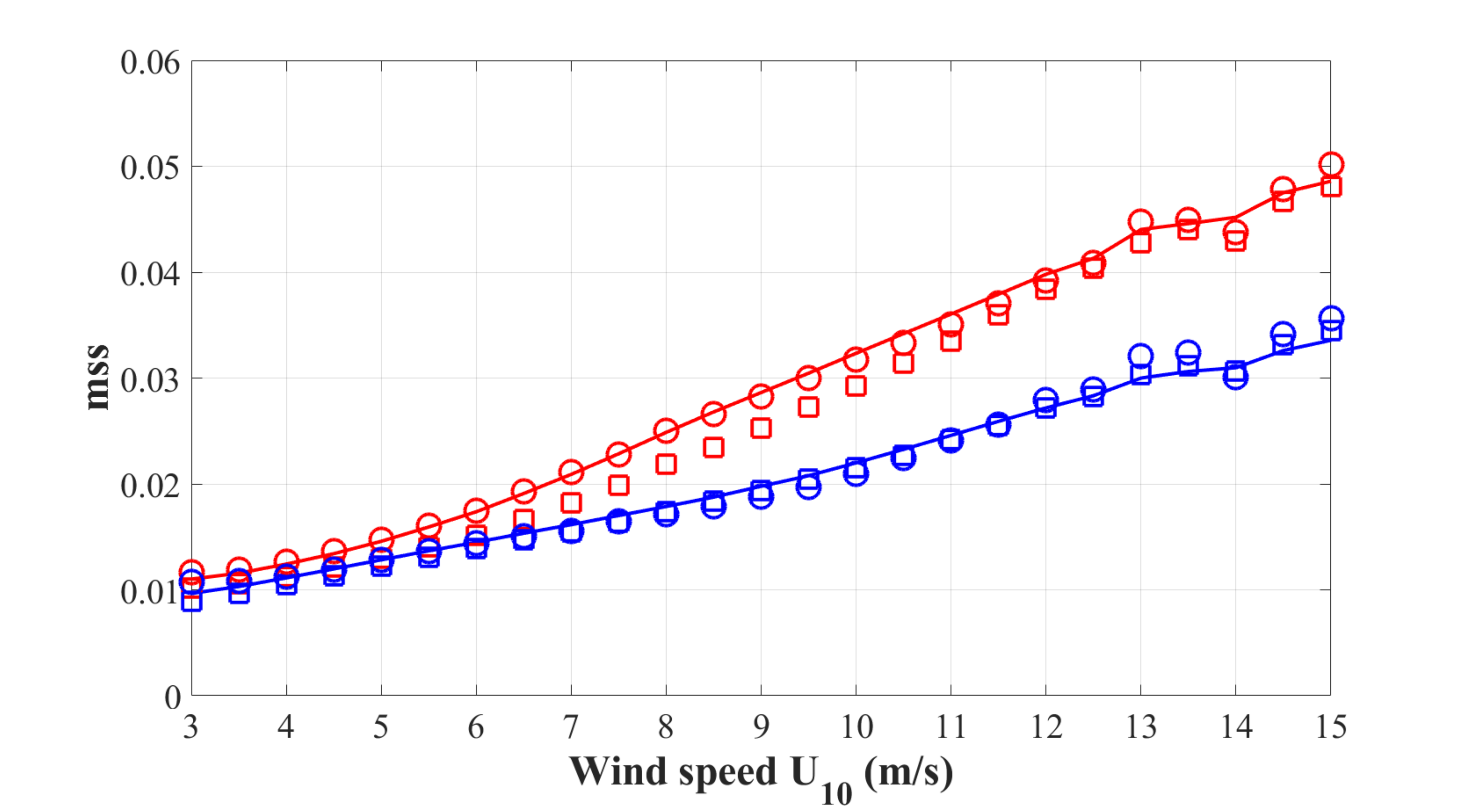}
  \caption{Comparison of the directional MSSs (solid lines) obtained with the IASI data using Eq. (\ref{calculemssfin}) and the MSS-shape's (squares) obtained with Eq. (\ref{defmssshape}). Also displayed (circles) are the MSSs inferred from the MSS-shape using  Eq. (\ref{mssshapetomss}), showing consistent results. Red and blue colors refer to the up- and cross-wind values, respectively.}
  \label{fig:mssshape}
\end{figure}

Finally, Fig. \ref{fig:mssshape} shows the evolution of the directional MSS-shape with wind speed and their comparison with the MSSs. Both are very close except in the upwind direction at moderate wind speed (5-10 m/s), where the MSS-shape is significantly smaller (by about $10\%$) than the upwind MSS . This is consistent with the evolution of the kurtosis coefficients (see Table \ref{tablemss})  which reach their largest values in the upwind direction and this range of wind speed. Recall that the peakedness of the distribution, which is quantified by the (excess) kurtosis coefficients $C_{40}$, $C_{04}$ and $C_{22}$, has been interpreted by \cite{chapron_JGR00} as the effect of compounding of the random processes influencing the sea-surface slopes through the variability of small-scale slope variance from one patch to another due to short-term natural processes such as wind gust or modulation by large wave groups. It thus seems that the maximal variability of the MSSs occurs at moderate wind speeds, which are also the most frequent. To check the consistency of the estimations we also display in Fig. \ref{fig:mssshape} the ``reconstructed'' MSS inferred from the MSS-shape through Eq. (\ref{mssshapetomss}). As can be seen, they are in excellent agreement with those estimated from the moments, except at the largest wind speeds where small differences appear, which indicates a reduced accuracy of either estimations.

\subsection{Comparisons with previous results}
\subsubsection*{The mean square slopes}
\begin{figure}\centering
 \includegraphics[scale=0.3]{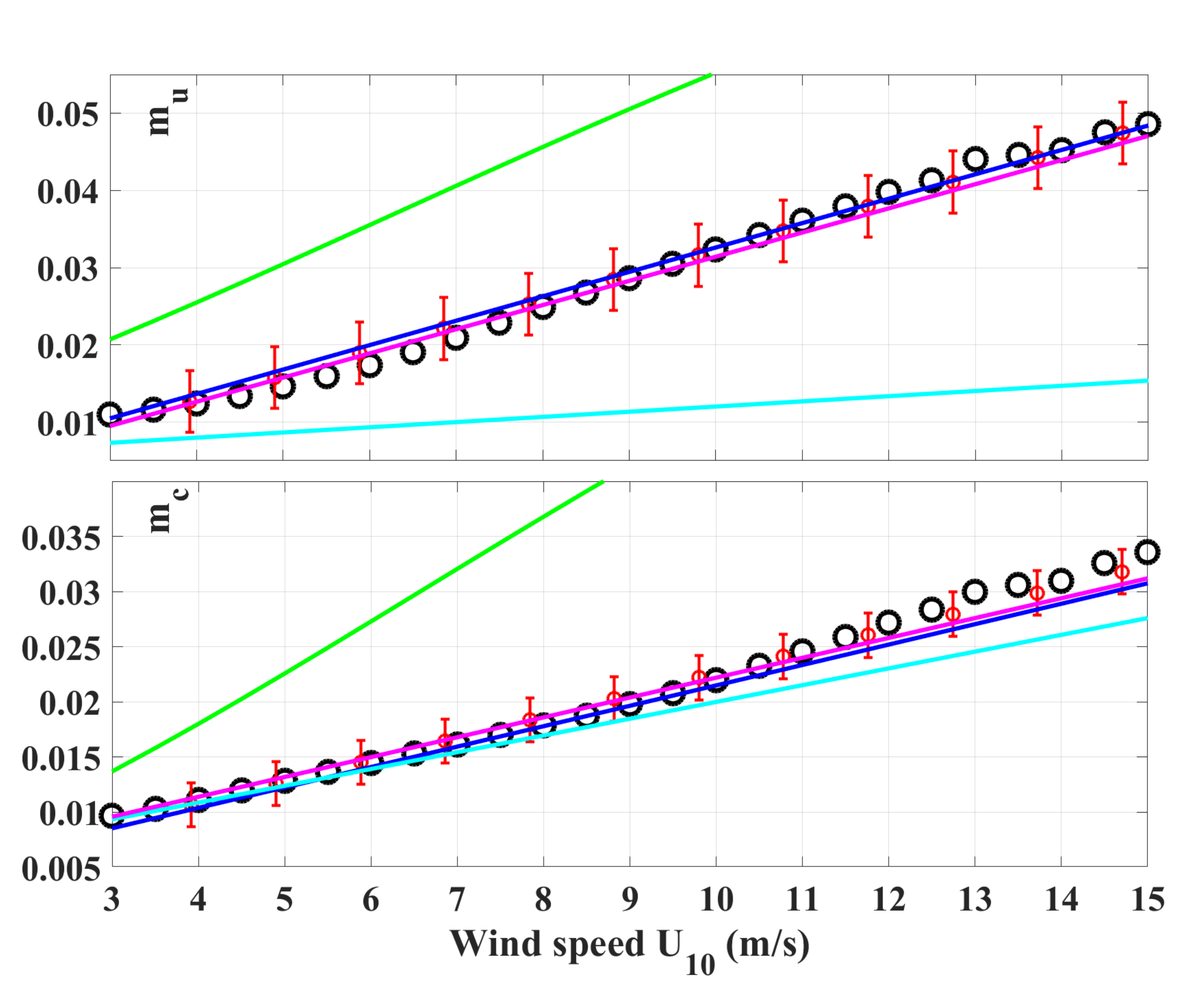}
  \caption{Evolution with wind speed of the up-  ($m_u$, left panel) and cross-wind ($m_c$, right panel) MSSs. The black circles represent our values (given in Table \ref{tablemss}), while the red lines correspond to the CM linear laws with associated error bars \citep{Cox54}. Also given are various other proposed parameterizations from the literature, by \cite{Mermelstein} (green), \cite{Ebuchi} (cyan), \cite{Breon} (blue) and \cite{Lenain} (magenta).}
  \label{fig:msswindnew}
\end{figure}

Figure \ref{fig:msswindnew} displays our $m_u$ and $m_c$ values together with some parameterizations versus wind speed available in the literature. Note that the wind speeds used by CM were measured at 12.5 m above the sea surface, while the other studies use an altitude of 10 m, and were thus converted at 10 m using a classical von Karman logarithmic profile, see e.g. \citep{tennekes73}, which resulted in their multiplication by $0.98$ for all wind speeds. As can be seen, our results confirm the anomalous behavior of those of \cite{Mermelstein} and \cite{Ebuchi} noticed by \cite{Zhang2010} and indicated by the comparisons with the other values \citep{Cox54,Breon,Lenain} in Figs. \ref{fig:msswindnew}. Furthermore, our values are fully compatible with the linear law of CM in Eq. (\ref{loiCM}), since they fall within the uncertainties on the latter. However, it must be reminded that the error bars given by CM \citep{Cox54} were obtained by calculating the overall RMS difference between their MSSs estimates at each available wind speed (in their case,  25 different values) and their associated linear fits, thus measuring the ability of the latter to describe the data rather than the uncertainties of their individual MSSs values. In contrast, our uncertainties are much smaller, thanks to the huge number of observations used, and enable to point out a small but statistically significant bias of the CM linear relationship for $m_u$ at moderate wind speeds (5-7 m/s), where our values are 5-8\% smaller. The possible origin of this small difference is further discussed in Sec. \ref{discussion}.

\subsubsection*{The kurtosis coefficients}
\begin{figure}\label{figkurtosis}\centering 
\includegraphics[scale=0.25]{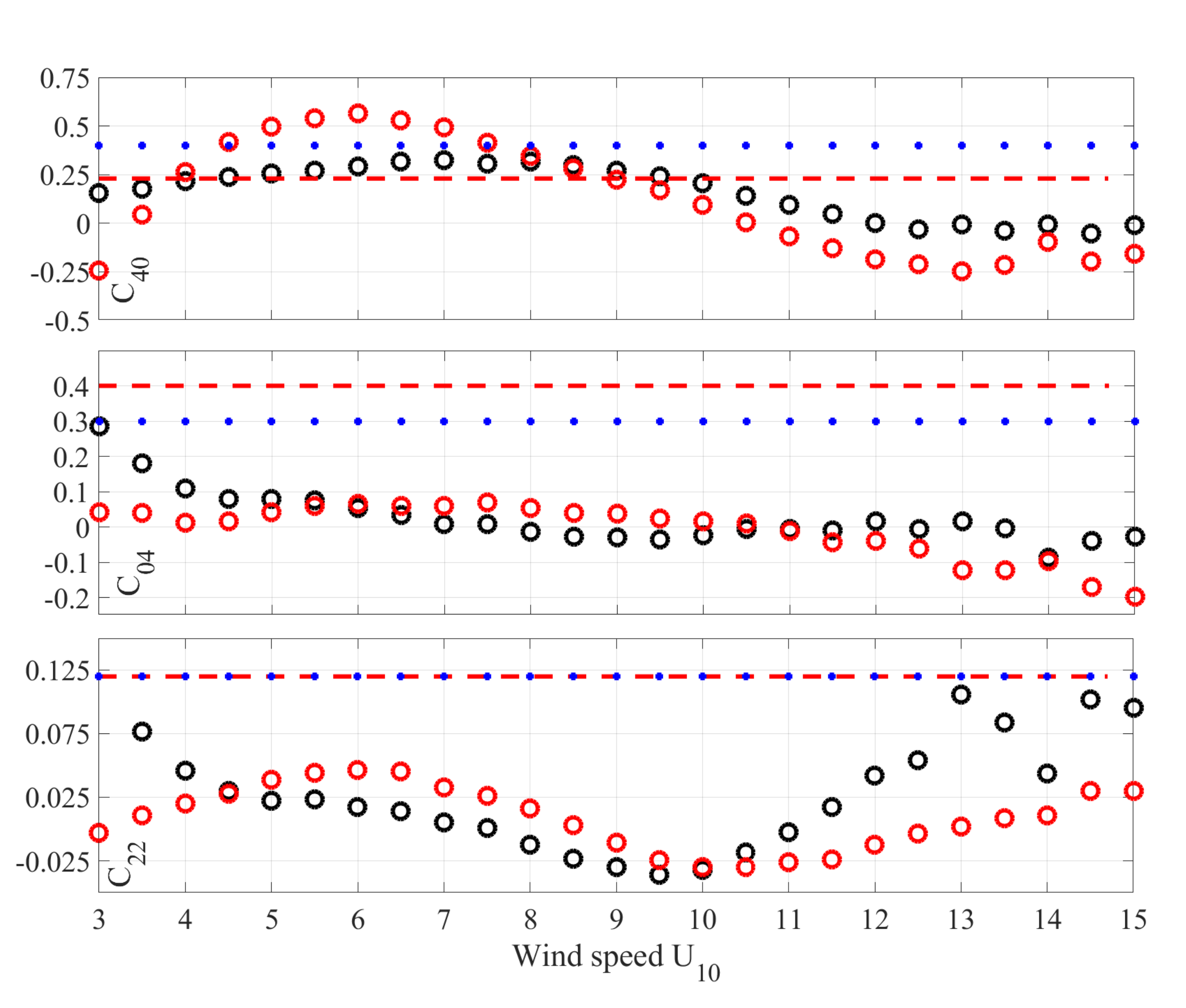}
  \caption{Evolution of the kurtosis GC coefficients with wind speed, as obtained from fits of the IASI data using the MSSs given by Eq. (\ref{mymss}) (black circles) or those of CM (red circles). The red dashed horizontal lines and the blue dots are the wind-independent values proposed by CM and BH.}
  \label{fig:kurtosiswind}
\end{figure}

Figure \ref{fig:kurtosiswind} shows the variations of the kurtosis coefficients with wind speed. The alongwind peakedness ($C_{40}$) ranges from $0$ to $0.32$ at moderate wind speed with maximal values reached between 6 and 9 m/s, slightly negative values being obtained beyond 12 m/s. It is thus smaller than the wind-independent (mean) values proposed by BH ($0.4\pm 0.1$) and CM ($0.23\pm 0.41$) but compatible with the latter when uncertainties are taken into account. Our $C_{22}$ and $C_{04}$ coefficients are very small (between $0$ and $0.1$) with slightly negative values for $C_{04}$ above 11 m/s. Our results here are thus significantly different from those of both CM ($C_{22}=0.12\pm 0.06,C_{04}=0.4\pm 0.23$) and BH ($C_{22}=0.12\pm 0.03, C_{04}=0.3\pm 0.05$). However note that CM could only obtain a very coarse estimation of the kurtosis coefficients based on fitting the parameters entering in the azimuthal expansion of Eq. (\ref{logfit}) with very large  uncertainties on the obtained results. Concerning BH, they found values of the same order as those of CM but, in contrast, their upwind kurtosis coefficient, $C_{40}$, is larger than their crosswind one, $C_{04}$, which is consistent with our results. Their uncertainties are small, but they were obtained by measuring the deviation of the fit to the data by wind-independent values and not from a proper error analysis. In addition, a closer inspection of the values in Fig. 5 of \cite{Breon} shows influences of wind speed in qualitative agreement with our results: $C_{40}$ shows a maximum at moderate wind speed with significantly lower values for both small and high wind speeds ; $C_{04}$ and $C_{22}$ remain almost constant at small and moderate wind speeds and their observed variations at higher wind-speed are less trustworthy due to the reduced number of observations. We were not able to evaluate the uncertainties on our determinations of the kurtosis coefficients since there is no simple way to measure the dispersion of values. However, the estimation of $C_{40}$, $C_{04}$, and $C_{22}$ that we made using a GC fit (See Sec. 5.2.2), is highly dependent on the assumed MSSs and small variations of the latter can induce large changes of the former. To evaluate the impact of the assumed input MSSs, we carried a second determination of the kurtosis coefficients by re-fitting the IASI data while imposing the MSSs values of CM instead of our owns. As shown by Fig. \ref{fig:kurtosiswind}, this leads to a much larger value of $C_{40}=0.6$ at 6 m/s, consistent with the findings of BH for this coefficient. However, no significant changes were observed for $C_{04}$ and $C_{22}$ which keep oscillating around zero. Note that slightly negative kurtosis coefficients at high wind speed were also found in the recent airborne experiments by \cite{Lenain}.

\subsubsection*{The skewness coefficients}
\begin{figure}\centering
 \includegraphics[scale=0.25]{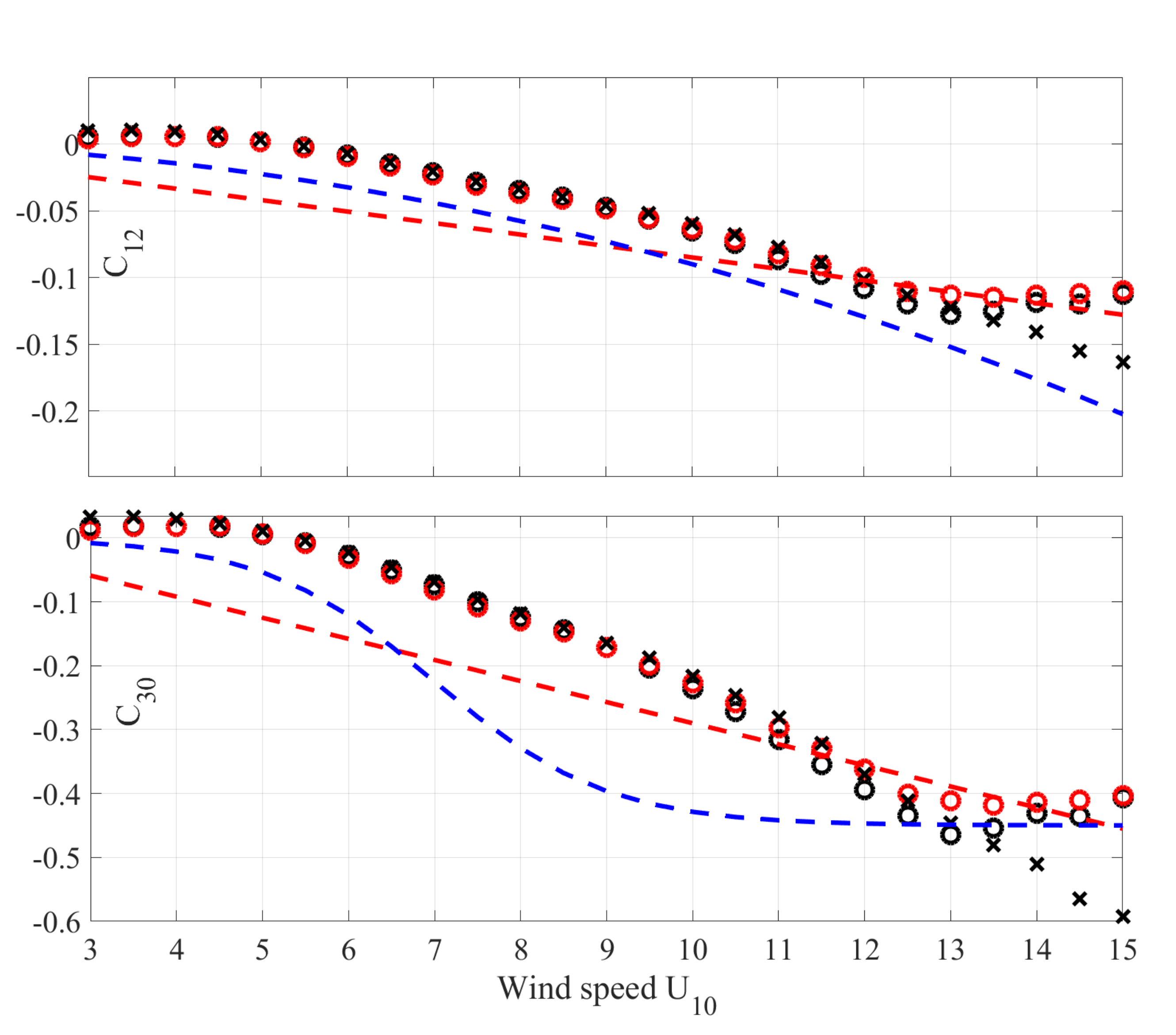}
  \caption{Evolution of the skewness GC coefficients with wind speed as fitted from the IASI data on the basis of the MSSs given by Eq. (\ref{mymss}) (black circles) or those of CM (red circles). Also shown are the coefficients obtained from the first moment formula of Eq. (\ref{C12moment})  (black crosses) and the values proposed by CM and BH (red and blue dashed lines).}
  \label{fig:skewcoeffs}
\end{figure}

Figure \ref{fig:skewcoeffs} shows the variations of the  skewness coefficients, in which we display the values obtained using the two available techniques (fit or moment) to check the consistency of our estimations. As can be seen, the two techniques lead to results that are in very good agreement, except beyond 13 m/s. This can be explained by the fact that the moment ${\cali M}_{11}$ cannot be estimated accurately for strong winds due to a slower decrease of $a_1$ with the wave tilt. This is demonstrated by computing the mean slopes, which should be zeros, i.e.:
  \be\label{meanslope}
  \begin{split}
 \int_{-\infty}^{+\infty}\int_{-\infty}^{+\infty} s_up(s_u,s_c)ds_uds_c= \pi {\cali M}_{21}=0.
 \end{split}
 \ee
 The numerical evaluation of ${\cali M}_{21}$ indeed  shows that it is of the order of $10^{-4}$ for slow winds and of $10^{-2}$ for the largest wind speed (15 m/s), a growing deviation from zero due to the increasing inaccuracy in the tail of the measured distribution. Now regarding the comparison of our skewness coefficients with those of CM and BH, Fig. \ref{fig:skewcoeffs} shows a good qualitative agreement, all results being negative and steadily decreasing with increasing wind speed. Furthermore, our results are fully compatible with those of CM when the uncertainties on their linear laws are taken into account. This also roughly stands for the difference between our results for $C_{12}$ and those of BH, but not for $C_{30}$, the BH given uncertainties on both coefficients being $\pm 0.01$. However, our results support the BH observation of a non-linear relationship with wind speed as opposed to CM. As for the kurtosis, we were not able to devise meaningful uncertainties for the skewness coefficients. However, as done before, we checked the influence of the MSSs by retrieving the skewness coefficients from fits of the IASI data using the CM MSSs as inputs. The results shown in Fig. \ref{fig:skewcoeffs} demonstrate a very good consistency between the values retrieved using the two sets of MSSs. 
 
\subsection{Discussion}\label{discussion}
 We here discuss, in view of the differences between our results and those of CM pointed out in the preceding section, the two main hypotheses made by \cite{Cox54}.
 
 The first is that they assumed that the 'complete' MSSs  can be obtained by applying a ``blanket increase'' to the 'incomplete' ones (i.e, those obtained by integration of the second moments up to the limiting slopes) regardless of the wind speed [see Eq. (\ref{renorm})]. This rule was established on the basis of the empirical fit using Eq. (\ref{logfit}), which was assumed to hold beyond the limit of the observed slopes,  that is $s_4=(4/\alpha_0')^{1/2}$, where $\alpha_0'$ is the quadratic slope coefficient appearing in Eq. (\ref{logfit}). In \cite{wentz_JGR76}, the variance renormalization procedure of CM was criticized on the ground that, based on the logarithmic coefficients [Eq. (\ref{logfit}), the upper bound $s_8=(8/\alpha_0')^{1/2}$ was not sufficient to reach convergence of the integral [Eq. (\ref{defsigma})] and was therefore arbitrary. We have checked this assumption by numerical simulations based on the GC model (not the logarithmic fit), and found that the integral Eq. (\ref{defsigma}) is well converged at $s_8$ and therefore the renormalization factor well-defined. However, contrarily to CM, we found the resulting factors $R_u/R_c$ to be wind-speed dependent, where we have defined:
\be
R_u=\frac{\int_{\abs s\leq s_8} s_u^2p(s_u,s_c)ds_uds_c}{\int_{\abs s\leq s_4} s_u^2p(s_u,s_c)ds_uds_c} , R_c=\frac{\int_{\abs s\leq s_8} s_c^2p(s_u,s_c)ds_uds_c}{\int_{\abs s\leq s_4} s_c^2p(s_u,s_c)ds_uds_c} .
\label{ratioR}
\ee
 
As shown by CM \citep{Cox56} by expanding the GC formula in power of the wave tilt, one has:  
  \be
  \label{alpha0}\alpha_0'=\frac{1}{4}\left(\frac{1}{m_u}+\frac{1}{m_c}\right)+\frac{1}{8}\left(\frac{C_{40}}{m_u}+C_{22}(\frac{1}{m_u}+\frac{1}{m_c})+\frac{C_{04}}{m_c}\right).
  \ee

\begin{figure}\centering
  \includegraphics[scale=0.25]{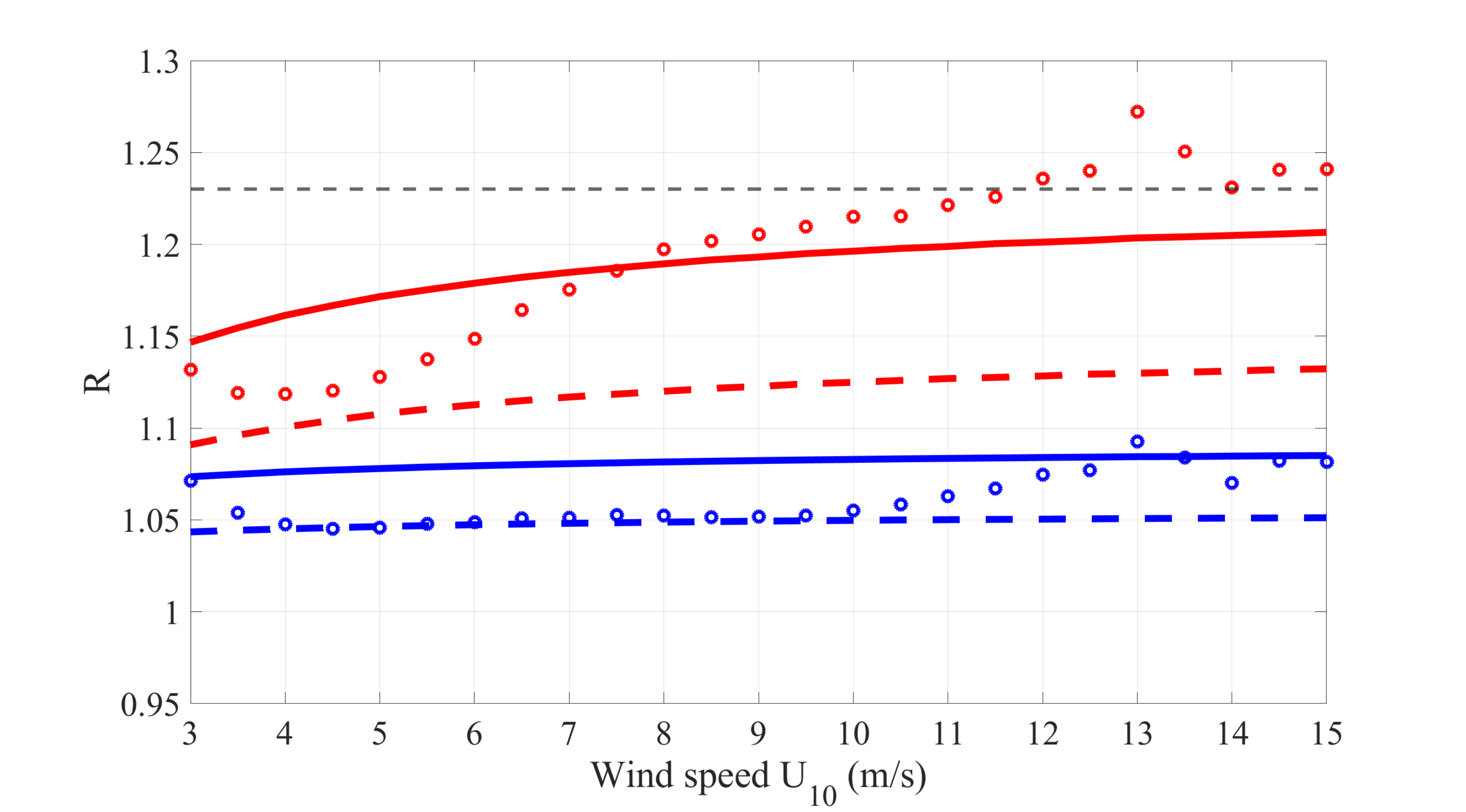}
  \caption{Evolutions of the 'complete to incomplete' variance MSSs ratios $R_u$ (in red)  and $R_c$ (in blue)  calculated using Eq. (\ref{ratioR}) and the datasets (see text) (D1), (D2) and (D3) represented by circles, dashed and full lines, respectively. The dashed black line denotes the constant factor $1.23$ assumed by CM, see Eq. (\ref{renorm})}
  \label{fig:ratiomss2}
\end{figure}

\begin{figure}\centering
  \includegraphics[scale=0.25]{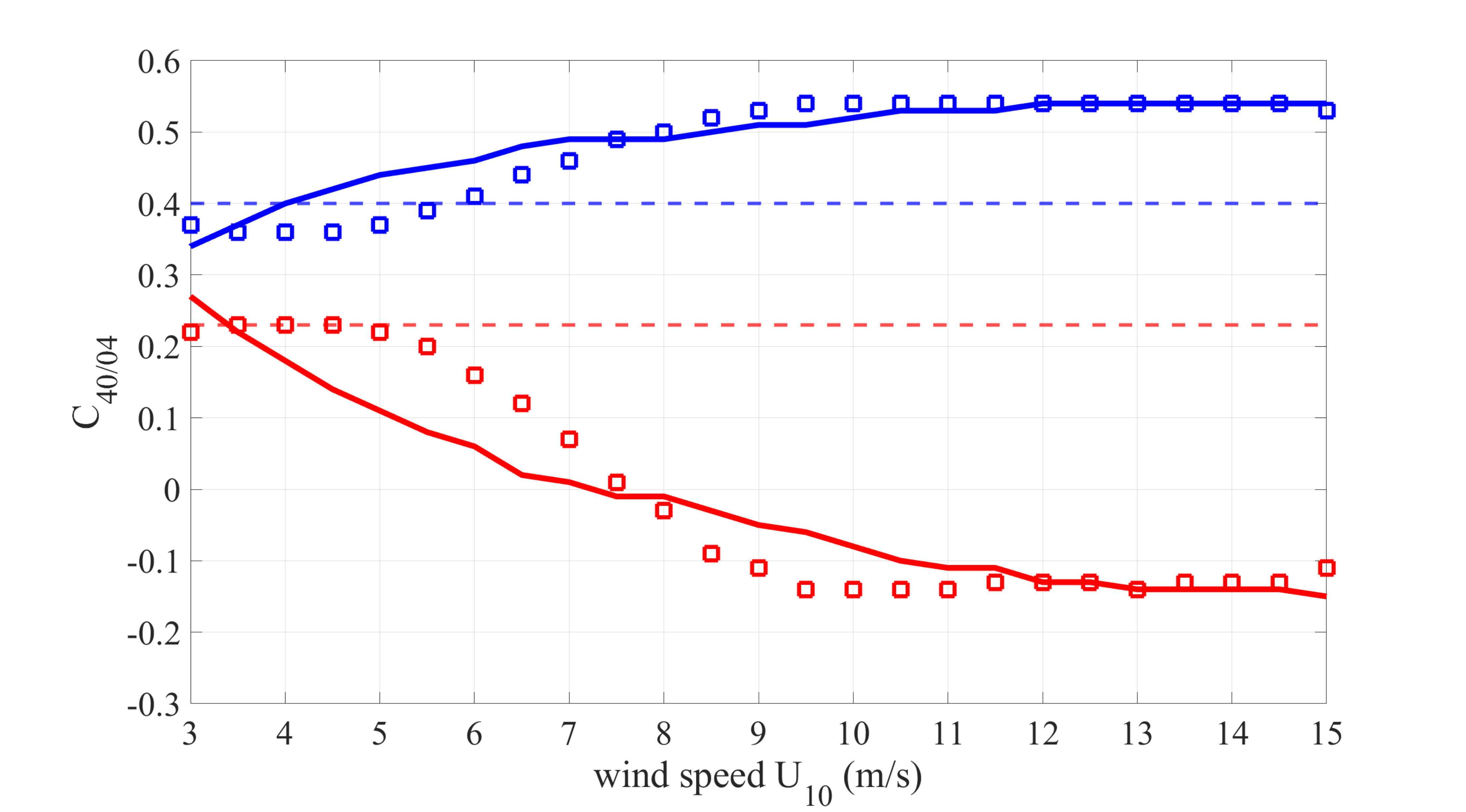}
  \caption{Upwind ($C_{40}$, in red) and cross-wind ($C_{04}$, in blue) kurtosis coefficients that best fulfill (see text) $R_u=1.23$ and $R_c=1.22$, assuming CM MSSs (solid lines) or IASI MSSs (squares). The dashed line denote the constant values of CM. }
  \label{fig:optimalkurt}
\end{figure}

  We then carried calculations of $R_u$ and $R_c$ using three different input datasets: (D1) The MSSs and kurtosis and skewness coefficients obtained (Table 1) from the  analysis of IASI data. (D2) The BH \citep{Breon} MSSs and the associated kurtosis and skewness coefficients. (D3) The CM \citep{Cox54} values of the MSSs and kurtosis and skewness coefficients. The resulting ratios $R_u$ and $R_c$ are shown in Fig. \ref{fig:ratiomss2}, calling for several remarks. First, it is clear that none of the input datasets leads to results consistent with the constant values $1.23/1.22$ mentioned by CM. Secondly, the ratio $R_u/R_c$ is very sensitive to the kurtosis coefficients, as demonstrated by the important differences between the results obtained using BH (D2) and CM (D3) parametrizations despite that the associated  MSSs are almost identical. In contrast, it is weakly dependent on the MSSs, as seen around 6 m/s where the CM and IASI MSSs differ most while the upwind kurtosis coefficients are close. If one assumes the correctness of the $1.23/1.22$ factors of CM, this means that these values can only be reached using accurate and wind-dependent values of the kurtosis coefficients corresponding to the specific sea states conditions encountered in the CM experiments. To elucidate this issue, we determined the kurtosis coefficients ($C_{40},C_{04}$) for which the computed values of $R_u$ and $R_c$ are 1.23 and 1.22. To reduce the number of unknowns we fixed the values of the MSSs and of $C_{22}$, which has the smallest uncertainty ($\pm 0.06$), to those of C.M The same exercise was carried out using the IASI MSSs. The results are shown in Fig. \ref{fig:optimalkurt} together with the mean values $C_{40}=0.23,\ C_{04}=0.4$ found by CM. It turns out that the kurtosis coefficients clearly vary with wind speed but have quite constant values at small (3-6 m/s) and large (9-15 m/s) wind speeds, where the mean values inferred by CM are respectively quite correct and very inaccurate. The use of IASI MSSs leads to results that are closer to the $C_{40/04}$ measured by CM, a fact that supports the small negative correction brought by IASI with respect to the CM linear law [Eq. (\ref{loiCM})] for the MSS in the upwind direction shown in Fig. \ref{fig:mssshape}. Another important outcome of this numerical test is the occurrence of a negative upwind kurtosis ($C_{40}\simeq -0.15$) at large wind speed, which tends to confirm the recent observations by \cite{Lenain}. Another specificity of the CM results is their very large crosswind kurtosis $C_{04}\simeq 0.55$ . This is clearly not representative of the average sea states observed with IASI that have a negligible crosswind kurtosis and a more pronounced upwind one. We hypothetize that this difference is specific to the wave conditions encountered by CM in Hawa\"i.


\textcolor{black}{The second important hypothesis made by CM is that of an homogeneous water surface with homogeneous sky illumination. For this, the presence of natural slicks has been neglected and the effect of cloud shadows were removed by manual selection of the image photographs. A special attention has also been given to the elimination of whitecaps, which correspond to dark spots on the photometric plates. These were removed from the measurements by careful visual inspection but the authors acknowledged that this was troublesome at large wind speeds. In their data analysis, BH also eliminated the contribution of foam and aerosols by rejecting the observations for which the off-glint reflectance was larger than some threshold. This was made possible by the multi-directional capability of the POLDER instrument, which could observe a same point from different viewing angles within a short fly time of the satellite. However, this is a very coarse criterion as the spatial resolution of the satellite data (a few tens of km$^2$) does not allow to resolve the foam patches on the sea surface. Hence eliminating the measurements with the highest off-glint reflectance does not exclude a contribution of whitecaps to the remaining measurements, albeit smaller. Furthermore, this technique can create a statistical bias in the slope distribution as it selects some specific sea states at given wind speed (those with less whitecaps). Nevertheless, despite an imperfect elimination of whitecaps in both the CM and BH studies, the retrieved MSSs are in excellent agreement, indicating a negligible impact of the foam leftover. }

\textcolor{black}{In our processing of the IASI measurement, cloudy and aerosol-contaminated observations have been removed from the data set, and natural surface slicks have been ignored. However, the configuration of the IASI instrument does not allow to measure glint and off-glint reflectance at the same ground location so that the white cap elimination procedure of BH could not be reproduced. Hence, the whitecap contribution, if any, is included in the IASI radiances but disregarded in our analysis for the retrieval of the wave-slope probability (and sea surface temperature). Now, evaluating this perturbation is an impossible task in the present state of knowledge. Indeed, due to the dispersion in the whitecap coverage models [see e.g. the review by \cite{Anguelova}], to the complexity of the geometrical and statistical description of breaking waves and foam [e.g. \citep{Sharkov}], to the opposite impacts on the radiometric parameters according to whether the foam is an active or passive stage \citep{Marmorino, Potter}, and to the increased contrast of sea foam reflectivity at large  incidence angle \citep{Branch,Niclos}, we have no reliable model at hand to correct precisely for the influence of whitecaps. }

\textcolor{black}{However, in the mid-infrared range ($3.6-4 \mu$m) used in the present study, the emissivity of sea water and foam have close values \citep{Branch}, a statement that also applies to the reflectance [respectively $\sim 2.5 \%$ and $\sim 3 \%$ at 4 $\mu$m, see Figure 5 of \cite{Salisbury1}]. Furthermore, the sea surface temperatures retrieved from IASI \citep{Capellenighttime,Capelledaytime} show an excellent agreement with in situ measurements [taking into account the cool-skin effect \citep{Fairall}] with an overall flat RMS difference when plotted against wind speed, suggesting that foam patches inside the field of view globally have a very  small influence on the brightness temperature around 4 $\mu$m. This weak radiometric contrast indicates that the presence of breaking waves have an effect on the sea surface radiance that is much smaller than in the visible and near-infrared region, where foam and seawater reflectivity can differ by one order of magnitude \citep{Frouin, Dierssen}. Given the small spatial coverage of whitecaps for the considered wind speeds (a few percents at 15 m/s wind speed, see e.g. \citep{Anguelova, Anguelova2}) this can only cause minor change in the measured surface radiance and therefore only impact the retrieval of very infrequent slopes. As mentioned in Section \ref{sec:IASI}, we believe that this partly explains the plateau observed in the retrieved  distribution at large slopes (see Figure \ref{fig:pdf7up}). The value of this offset defines the maximal available slope that can be obtained through the Geometrical Optics approach. Recall that this maximal slope was found larger by a few degrees than CM maximal slope, a finding which is consistent with the reduced reflectivity of whitecaps (hence a lower plateau).
This bias has been corrected by subtracting a constant value that allows to recover the normalization condition of the pdf and eliminate the mean whitecap contribution. Even though this is a coarse correction, it is sufficient for an accurate determination of the MSSs as those are evaluated through the lowest moments of the distribution (which converge rapidly and have little sensitivity to the tail). We therefore believe that there is no significant contamination of the MSSs by whitecaps in our data processing. This explains the good agreement between our values and those of CM, except perhaps for the largest wind speeds (> 13 m/s). In their study using an airborne lidar on the ocean and a laser glint counting technique, \citep{Lenain} also reached, despite the fact that they did not filter out the whitecaps, a good agreement with CM and BH for the retrieved MSSs. Their skewness and kurtosis coefficients, however, were very different from those of CM and BH and the authors hypothetized that this might be due to the presence of whitecaps and breaking. Since these coefficients were evaluated with a direct calculation of the third and fourth moments of the distribution, which are very sensitive to the tail of the distribution, their hypothesis is consistent with our analysis. In our case, however, the higher-order Gram-Charlier parameters were obtained through a least-square fitting of the symmetric and asymmetric part of the distribution (once the MSSs obtained), a technique that prevents from computing higher moments of the distribution and gives more weight to the central part of the pdf. With this procedure, the accuracy of the retrieved skewness and kurtosis is mainly conditioned by the accuracy of the MSSs since any error in the latter can be partially compensated by changing the kurtosis in the best fit.}

\begin{figure}
   \centering
  \includegraphics[scale=0.25]{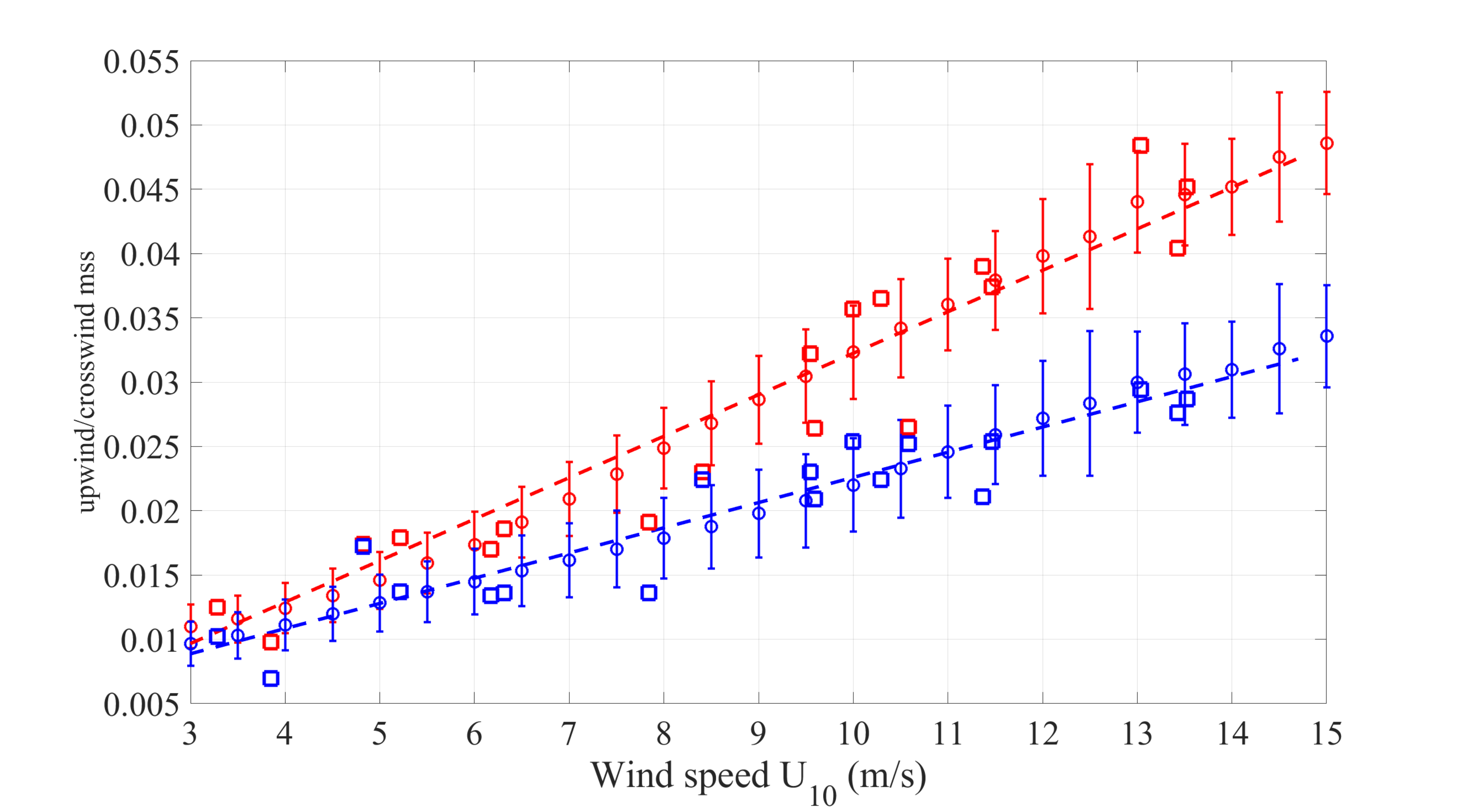}
  \caption{Original MSSs data derived by CM \citep{Cox56} (squares), the linear laws deduced from the latter (dashed lines)  and comparison with those (circles) derived in the present work (Table \ref{tablemss}), the error bars indicating the range of variability. The blue and red colors refer to $m_c$ and $m_u$, respectively.}
  \label{fig:msswind_dataCM}
   \end{figure}

The third assumption implicitly made by CM to obtain the laws in Eq. (\ref{loiCM}), which is a major one, is that the wave-slope probabilities are only driven by the speed and direction of the wind. Together with most previous studies, including those of \cite{Breon} and \cite{Ebuchi}, our analysis also used this hypothesis. However recall an important difference, useful for the discussion made below, between the latter three investigations and that of CM:  In the former, the MSSs (and the other parameters of the GC expansion) were obtained, for each wind speed, from very large numbers of observations, while only one photograph was taken for each wind condition by CM. This implies that, if a parameter other than the wind also monitors the wave-slope probabilities, its influence in the retrieved pdf for a given wind speed has been largely reduced by the averaging made in studies based on large observation datasets, which is not the case for the CM study. Now, recall that the measured probabilities show, from one observation to another, an important scatter around the fitted pdf (see Figs. \ref{fig:pdf7up} and \ref{fig:exemplefit}). Note that comparable scatters were also obtained in the other studies involving great numbers of data [see the error bars in the Figs. 3 of \citep{Breon} and \citep{Ebuchi}]. In our case, this dispersion could be due to several effects, including: (E1) the noise on the IASI radiances and random errors in the forward model used to determine (see Sec. 3) the probabilities; (E2) inaccuracies in the assumed wind speeds and directions, which were provided by the ECMWF/ERA5 reanalysis (see the first paragraph of Sec. 4) ; (E3) the influence of meteo-oceanographic variables other than the wind on the wave-slope statistics. A careful analysis of our results and procedures shows that the first explanation, (E1), can be ruled out, at least for the large values of the retrieved probabilities, typically for $p(s_u,s_c,U)$ above 1, which are accurately determined. Concerning the second explanation, (E2), an error analysis shows that, in many cases, the observed scatter of the probabilities is by far beyond what could result from the uncertainties on the wind speed and direction. This is particularly the case near the inflection point where $\partial p(s_u,s_c,U)/ \partial U=0$, since the probability is here practically insensitive to the wind but close to $p(0,0,U)/e$ so that it is significant and thus well determined. It thus appears that the third explanation, (E3), stands in many cases, and the influences of the significant wave height, Richardson number, fetch, wave age, \textit{etc}, may be invoked. From this point of view, recall that some previous studies point out the roles played by the wave height \citep{SWH1,SWH2,stopaOM2016}, the swell \citep{hwang_JGR08}, and the degree of atmospheric stability \citep{Hwang,shaw}. However note that the findings of these last two references were not confirmed by \cite{Ross} and \cite{Lenain} and that the relative impact of swell on the MSS is still subject to some controversy \cite{hwang_JGR09comment,hauser_JGR09reply}. Clarifying the contribution of each of the above mentioned factors requires a substantial analysis which goes beyond the scope of this paper. However, in order to quantify the effect of the scatter observed in the present work on the individual values of the MSSs, we used an isotropic approximation of the probability at the origin $s_u=s_c=0$, $p_0\simeq 1/\pi m_t)$ (with $m_t\simeq 2 m_u\simeq 2 m_c$), from which we obtain $\Delta m_{u/c}\simeq \pi m_{u/c}^2 \Delta p$, where $\Delta p$ is the standard deviation of the measured probabilities around  $s=0$ (for $s<\epsilon$, where $\epsilon$ is taken to be a small fraction of the RMS slope). This leads to values of $\Delta m_{u/c}/m_{u/c}$ that decrease with wind speed and vary between typically 16 \% and 10 \%. It is important to here emphasize that $\Delta m_{u/c}$ quantifies the potential variability of the MSSs obtained from one individual observation to another and \textit{not} the uncertainty on the values given in Table \ref{tablemss}. Indeed, the latter are considerably more accurate, since they were derived from huge numbers of data points, as indicated by Fig. \ref{fig:pdf7up}. It is now interesting to compare our MSSs with the CM original data for a clean sea listed in Table I of \cite{Cox56}. This is done in Fig. \ref{fig:msswind_dataCM} where the vertical bars around our values indicate the range $\Delta m_{u/c}$ of variability of $m_{u/c}$ associated with the above discussed influence of meteo-oceanographic variables other than the wind. As can be seen, the CM values full within this range so that their unsmooth evolution with wind speed and their local differences with our results can be explained by the fact that the wind over the observed sea spot may not be the only parameter driving the wave-slope pdf. This reconciles the two datasets and confirms their compatibility.

\section{Conclusion}
We have determined the statistics of the slopes of the wave facets from space-borne observations of the sea surface. This was done using a proper treatment of the information brought by the contribution of solar photons to the radiances collected, around 3.8 $\mu$m, by the Infrared Atmospheric Sounder Interferometer (IASI). The obtained dataset was analyzed using an original and robust approach providing the wind-dependent 7 parameters of a Gram-Charlier description of the wave-slopes probability distribution function. The results are fully consistent, when uncertainties are considered, with those of the pioneer study of  CM \citep{Cox54}, which so far remains the reference. However, thanks to the unprecedented number (about 150 million) of IASI-retrieved probabilities and to the excellent absolute calibration of the instrument, we were able to accurately quantify  the influence of wind speed, not only on the mean square slopes but also on the kurtosis and skewness parameters, which  CM could not do. We pointed out a slightly reduced upwind MSS with respect to the CM linear law with wind speed, a difference for which we provided two possible explanations: The inadequacy of the linear fit in wind speed or specific sea states in the CM experiments that differ from the mean  conditions encountered in the present study. The kurtosis and skewness coefficients were shown to follow a clear tendency with wind speed, in qualitative but not quantitative agreement with the last determinations by BH. The dominant kurtosis coefficient was found upwind ($C_{40}$) with maximal values ($\simeq 0.3$) reached at moderate wind speeds while the other 2 coefficients ($C_{04},C_{22}$) were found quite constant and close to zero. Non-linear wind-speed dependences were obtained for the skewness coefficients but their absolute values were found smaller than those of BH.

We also pointed out and unexpected and so far undetected change of the sign of the up/down wind asymmetry with growing wind speed that we cannot explain but which certainly reveals some refined effects in the mechanisms governing the wave facets creation.

Finally, the IASI-retrieved wave-slope probabilities for given values of the wind speed and along- and cross-wind slopes show a significant scatter. A preliminary analysis indicates that the latter is often beyond the uncertainties on our determinations and may thus confirm that wind is not the only variable monitoring the wave-slope pdf. Investigating this key issue  and  quantifying the influences of other oceanic and atmospheric variables (the significant wave height, the atmospheric stability, fetch, \textit{etc}) is beyond the scope of the present paper but it will be the subject of a forthcoming investigation.

\section*{Declaration of Competing Interest}
The authors declare that they have no known competing financial interests or personal relationships that could have appeared to influence the work reported in this paper.

\section*{Acknowledgment}
This work has been supported in part by CNRS, CNES and Ecole polytechnique. VC and JMH  acknowledge the IPSL mesocenter ESPRI facility for computer simulations, as well as EUMETSAT and the Aeris infrastructure (\href{https://www.aeris-data.fr/}{Aeris}) for  access to the IASI Level 1c data. 

\section*{Appendix: Forward model equations}
Note that more details can be found in \cite{Capelledaytime}, and that the equations given below are written for the geometry of Fig. \ref{fig:IASI}.
\paragraph{The surface-emission contribution I$_{surf}$} Assuming that the sea-surface directional emissivity $\epsilon(\sigma,\theta_I)$   does not depend on the SST ($T_{surf})$, and since both $\epsilon(\sigma,\theta_I)$ and the black-body radiance $I_{BB}(\sigma,T_{surf})$ are constant over the narrow instrument function $F_{I}(\Delta\sigma)$ of IASI, the first term in Eq. (\ref{eqIASI}) is given by:
\begin{equation}
I_{surf}(\sigma,\theta_I,T_{surf})=\epsilon(\sigma,\theta_I)\tau(\sigma,\theta_I)I_{BB}(\sigma,T_{surf}) ,
    \label{eqIsurf}
\end{equation}
with
\begin{equation}
\tau(\sigma,\theta_I)=\int_{-\infty}^{+\infty}F_{I}(\sigma-\sigma')exp\left[-\int_{0}^{z_{Inst}}\frac{\alpha(\sigma',z)}{\cos(\theta_I)}dz\right]d\sigma' .
     \label{eqTau}
\end{equation}
where $z_{Inst}$ is the altitude of IASI, and $\alpha(\sigma,z)$ is the atmospheric absorption coefficient at wave number $\sigma$ and  altitude z. 

\paragraph{The upward atmospheric-emission contribution $I_{atm}^+$} Using the same notations, the second term in Eq. (\ref{eqIASI}), representing the emission of the atmospheric column between the surface and the instrument, is given by:
\begin{equation}
\begin{split}
    &I_{atm}^+(\sigma,\theta_I)=\int_{-\infty}^{+\infty}F_{I}(\sigma-\sigma') \int_{0}^{z_{Inst}}\frac{I_{Loc}(\sigma',z,\theta_I)}{\cos(\theta_I)}dzd\sigma' ,
        \label{eqIup}
        \end{split}
\end{equation}
with
\begin{equation}
\begin{split}
 I_{Loc}(\sigma,z,\theta_I)&=I_{BB}\left[\sigma,T(z)\right]\alpha(\sigma,z) \exp\left[-\int_{z}^{z_{Inst}}\frac{\alpha(\sigma,z')}{\cos(\theta_I)}dz'\right] .
    \end{split}
    \label{eqIloc}
\end{equation}

\paragraph{The downward atmospheric-emission contribution $I_{atm}^-$} The third term in Eq. (\ref{eqIASI}) is the emission of the atmosphere toward the surface that is then reflected and transmitted up to the instrument. Its rigorous calculation is computationally costly since it requires an integration over all possible downward-emission directions, each weighted by the proper  bi-directional reflectivity. However, as demonstrated in \cite{Capelledaytime}, $I_{atm}^-$, which makes a small contribution to the total radiance, can be  precisely evaluated by considering a single direction with a zenith angle $\theta_{down}$= 53$^{\circ}$. One thus has:
\begin{equation}
\begin{split}
 I_{atm}^-(\sigma,\theta_I)&=\int_{-\infty}^{+\infty}F_{I}(\sigma-\sigma') \int_{0}^{+\infty}I_{Loc'}(\sigma',z,\theta_{down})\\
 &dz/\cos(\theta_{down})d\sigma', 
 \end{split}
        \label{eqIdown}
\end{equation}
with
\begin{equation}
\begin{split}
 &I_{Loc{'}}(\sigma,z,\theta_{down})=\left[1-\epsilon(\sigma,\theta_{down})\right]exp\left[-\int_{0}^{z_{Inst}}\frac{\alpha(\sigma,z{'})dz'}{\cos(\theta_I)}\right]\\
 &\times I_{BB}\left[\sigma,T(z)\right]\alpha(\sigma,z)exp\left[-\int_{0}^{z}\frac{\alpha(\sigma,z')}{\cos(\theta_{down})}dz'\right] .
 \end{split}
\label{eqIlocp}
\end{equation}

\paragraph{The solar contribution I$_{sun}$} Finally, the contribution of the reflected solar photons, denoted as $I_{sun}$ in Eq. (\ref{eqIASI}), is given by:
\begin{equation}
\begin{split}
        &I_{sun}(\sigma,\theta_I,\varphi_I,\theta_S,\varphi_S)=\frac{\rho(\sigma,\theta_I,\varphi_I,\theta_S,\varphi_S)d\Omega_S}{4\cos^4(\theta_W)}\\
        &\times \int_{-\infty}^{+\infty}F_{I}(\sigma-\sigma')I^0_{Sun}(\sigma')\exp\left[-\int_{0}^{z_{Inst}}\frac{\alpha(\sigma',z)}{\cos(\theta_I)}dz\right]\\
        &\times \exp\left[-\int_{0}^{+\infty}\frac{\alpha(\sigma',z)}{\cos(\theta_S)}dz\right]d\sigma',
    \end{split}
     \label{eqIsun}
\end{equation}
where $I_{sun}^0$$(\sigma)$, d$\Omega_S$, $\theta_{w}$, and $\rho$(…) respectively denote the (disk-averaged) extraterrestrial radiance emitted by the sun, the solid angle within which the sun is seen from Earth, the wave-tilt angle, and the bi-directional spectral reflectivity of the surface. These last two quantities are given by:
\begin{equation}
tg(\theta_w)^2=\frac{\sin(\theta_{S})^2+\sin(\theta_{I})^2+2\sin(\theta_{S})\sin(\theta_{I})\cos(\varphi_{S}-\varphi_{I})}{\left[\cos(\theta_{S})+\cos(\theta_{I})\right]^2},
\end{equation}
and 
\begin{equation}
\rho(\sigma,\theta_I,\varphi_I,\theta_S,\varphi_S)=\rho(\sigma,\theta_{Inc})=\frac{R_{\parallel}^2(\theta_{Inc})+R_{\perp}^2(\theta_{Inc})}{2} ,
\label{eqRo}
\end{equation}
with (assuming a unity value of the index of refraction of air)
\begin{equation}
R_{\parallel}(\theta_{Inc})=-\frac{n_{SW}\cos(\theta_{Inc})-\cos(\theta_{Trans})}{n_{SW}\cos(\theta_{Inc})+\cos(\theta_{Trans})},
\end{equation}
and
\begin{equation}
R_{\perp}(\theta_{Inc})=+\frac{\cos(\theta_{Inc})-n_{SW}\cos(\theta_{Trans})}{\cos(\theta_{Inc})+n_{SW}\cos(\theta_{Trans})},
\end{equation}
where $n_{SW}(\sigma)$ denotes the complex refractive index of sea water and the incidence and transmission angles are given by:

\begin{equation}
\cos(\theta_{Inc})=\sqrt{\frac{1+\sin(\theta_{S})\sin(\theta_{I})\cos(\varphi_{S}-\varphi_{I})+\cos(\theta_{S})\cos(\theta_{I})}{2}}.
\label{eqthetetrans}
\end{equation}

\begin{equation}
\sin(\theta_{Trans})=\sin(\theta_{Inc})/n_{SW}(\sigma).
\end{equation}

\end{document}